\documentclass[aps, prx, superscriptaddress, twocolumn, longbibliography, nofootinbib]{revtex4-1}

\usepackage{amssymb}
\usepackage{amsmath}
\usepackage{graphicx}
\usepackage{amsfonts}
\usepackage[usenames, dvipsnames]{xcolor}
\usepackage{bm}
\usepackage{bbm}
\usepackage{color}
\usepackage{hyperref}
\hypersetup{pdfnewwindow=true,
            colorlinks=true,
            linkcolor=blue,
            citecolor=blue,
            urlcolor=blue}
\usepackage{comment}
\usepackage{soul}
\usepackage[normalem]{ulem}
\usepackage[caption=false, justification=centerlast]{subfig}

\DeclareMathOperator{\Tr}{Tr}
\DeclareMathOperator{\var}{var}
\DeclareMathOperator{\erf}{erf}

\newcommand{\bra}[1]{\langle #1|}
\newcommand{\ket}[1]{|#1\rangle}

\newcommand{\rom}[1]{\uppercase\expandafter{\romannumeral #1\relax}}

\newcommand{\av}[1]{\langle #1 \rangle}

\newcommand{\C}{\mathcal{C}}
\newcommand{\F}{\mathcal{F}}
\newcommand{\M}{\mathcal{M}}
\newcommand{\N}{\mathcal{N}}

\newcommand{\bea}{\begin{eqnarray}}
\newcommand{\eea}{\end{eqnarray}}
\newcommand{\beaa}{\begin{eqnarray}\begin{aligned}}
\newcommand{\eeaa}{\end{aligned}\end{eqnarray}}

\newcommand{\gtl}[1]{{\color{black}#1}}

\newcommand{\kar}[1]{\textcolor{black}{#1}}
\newcommand{\jon}[1]{\textcolor{black}{#1}}
\newcommand{\mar}[1]{\textcolor{black}{ #1}}

\newtheorem{theorem}{Theorem}

\begin{document}

\title{Optimal Quantum Thermometry with Coarse-grained Measurements}

\author{Karen V. Hovhannisyan\textsuperscript{\#}}
    \email{khovhann@ictp.it}
    \affiliation{The Abdus Salam International Centre for Theoretical Physics (ICTP), Strada Costiera 11, 34151 Trieste, Italy}
    
\author{Mathias R. J{\o}rgensen\textsuperscript{\#}}
    \email{matrj@fysik.dtu.dk}
    \affiliation{Department of Physics, Technical University of Denmark, 2800 Kongens Lyngby, Denmark}
    
\author{Gabriel T. Landi\textsuperscript{\#}}
\email{gtlandi@if.usp.br}
    \affiliation{Instituto de F\'{i}sica da Universidade de S\~{a}o Paulo,  05314-970 S\~{a}o Paulo, Brazil.}
    
\author{\'{A}lvaro M. Alhambra\textsuperscript{\#}}
    \email{alvaro.alhambra@mpq.mpg.de}
    \affiliation{Max-Planck-Institut fur Quantenoptik, D-85748 Garching, Germany}
    
\author{Jonatan B. Brask}
\email{jonatan.brask@fysik.dtu.dk}
    \affiliation{Department of Physics, Technical University of Denmark, 2800 Kongens Lyngby, Denmark}
    
\author{Mart\'{i} Perarnau-Llobet}
\email{marti.perarnaullobet@unige.ch}
    \affiliation{D\'{e}partement de Physique Appliqu\'{e}e, Universit\'{e} de Gen\`{e}ve, 1211 Geneva, Switzerland}

\begin{abstract}
Precise thermometry for quantum systems is important to the development of new technology, and understanding the ultimate limits to precision presents a fundamental challenge. It is well known that optimal thermometry requires projective measurements of the total energy of the sample. However, this is infeasible in even moderately-sized systems, where realistic energy measurements will necessarily involve some coarse graining. Here, we explore the precision limits for temperature estimation when only coarse-grained measurements are available. Utilizing tools from signal processing, we derive the structure of optimal coarse-grained measurements and find that good temperature estimates can generally be attained even with a small number of outcomes. We apply our results to many-body systems and nonequilibrium thermometry. For the former, we focus on interacting spin lattices, both at and away from criticality, and find that the Fisher-information scaling with system size is unchanged after coarse-graining. For the latter, we consider a probe of given dimension interacting with the sample, followed by a measurement of the probe. We derive an upper bound on arbitrary, nonequilibrium strategies for such probe-based thermometry and illustrate it for thermometry on a Bose-Einstein condensate using an atomic quantum-dot probe.
\end{abstract}

\maketitle


\section{Introduction}

\def\thefootnote{\#}\footnotetext{These authors contributed equally to this work.} \def\thefootnote{\arabic{footnote}}\setcounter{footnote}{0}

Thermometry is a basic metrological task, vital throughout science and technology. Estimating temperature is important on all scales, ranging from astronomical bodies with temperatures in the millions of Kelvin to atomic systems near absolute zero. In particular, applications of thermometry in nano- or micro-scaled devices are becoming increasingly relevant as technology advances \cite{Giazotto_2006, Yue_2012, Mehboudi_2019rev, Carlos_2016}. Examples include, for instance, accurate temperature estimation of ultra-cold gases \cite{Marzolino_2013, Sabin_2014, Onofrio_2016, Mehboudi_2019, Bouton_2020}, in electronic systems \cite{Gasparinetti_2015, Halbertal_2016, Karimi_2020}, or the use of atomic-size devices, such as colour centers in diamond or quantum dots, as probes to be employed in a variety of systems \cite{Higgins_2013, Kucsko_2013, Neumann_2013, Haupt_2014}. At these scales, quantum effects have significant influence on the achievable precision. It is therefore important to understand what the fundamental limits for temperature estimation in quantum systems are.

Quantum features offer both advantages and challenges to thermometry \cite{Pasquale_2018, Mehboudi_2019rev}. Advantages range from measurement enhancements due to strong coupling \cite{Correa_2017, Hovhannisyan_2018, Mehboudi_2019}, correlated probes \cite{Seah_2019, Planella_2020}, or nonequilibrium probes \cite{Brunelli_2011, Brunelli_2012, Jevtic_2015, Pasquale_2017, Cavina_2018, Mitchison_2020, Mancino_2020, Henao_2020}. The challenges are related to the inherent difficulty of accessing information in quantum systems, due, for instance, to measurement backaction or natural limitations in performing high-resolution measurements \cite{Potts_2019, Jorgensen_2020}.

\begin{figure}
    \centering
    \includegraphics[width=0.5\textwidth]{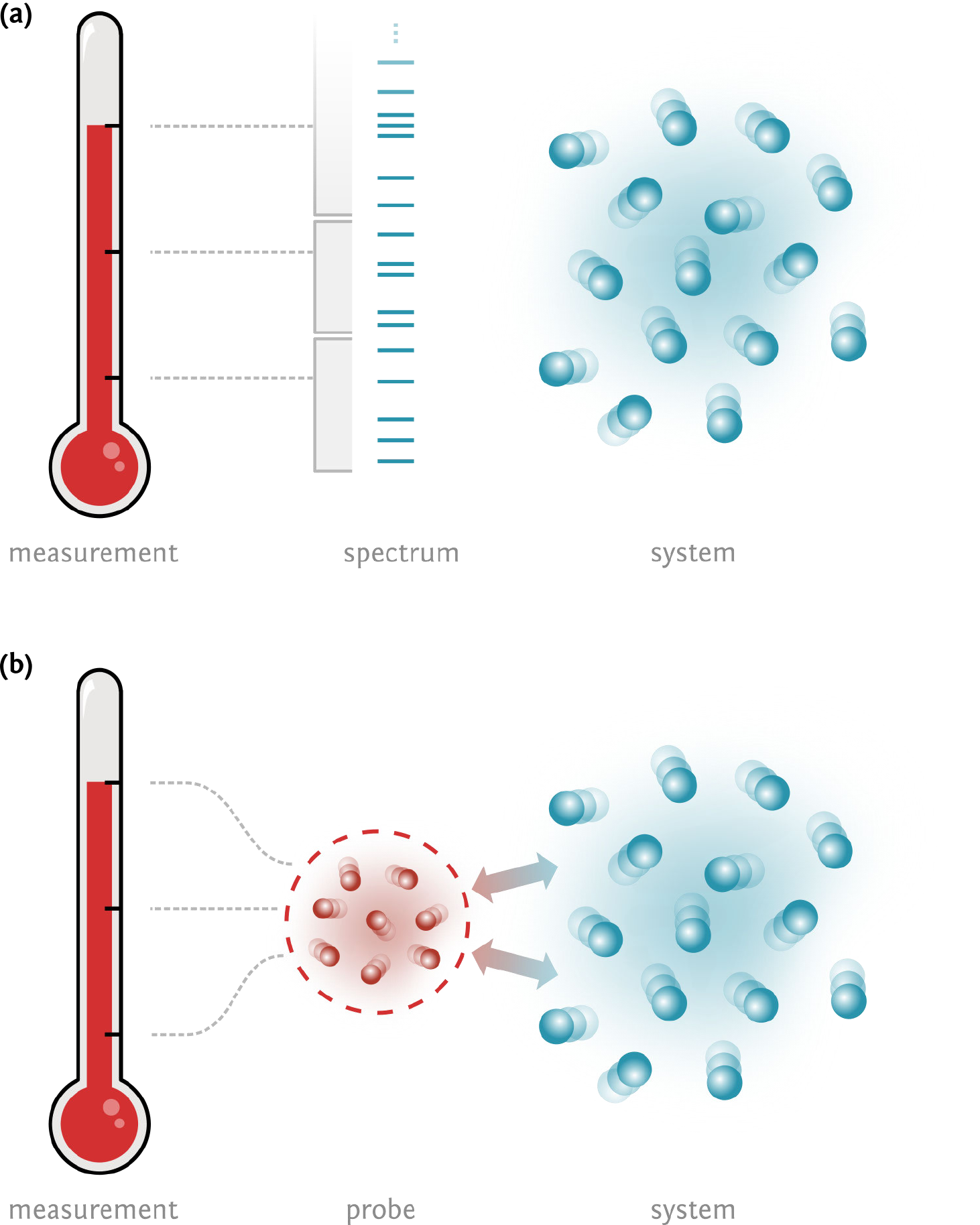}
    \caption{\textbf{(a)} Thermometry with coarse-grained energy measurements. The measurement can be understood as resulting from post-processing of a fine-grained, projective energy measurement. Energies are grouped into bins and a single outcome is assigned to each bin. \textbf{(b)} Thermometry with probe-based measurements. A probe interacts unitarily with a target system. A measurement is then performed on the probe alone and used to infer the temperature of the target.}
    \label{fig:coarsegraining}
\end{figure}

When the measurement resolution is unlimited, the ultimate precision of temperature estimation allowed by quantum mechanics is obtained by performing projective measurements of energy~\cite{Correa_2015, Paris_2015, Campbell_2018}. However, for large (or even moderately-sized) many-body systems, one seldom has access to measurements which distinguish individual energy levels. Instead, one usually 
\kar{measures only a local subsystem of the sample \cite{Pasquale_2016, Palma_2017, Hovhannisyan_2018} or performs a  global measurement with only a finite resolution \cite{Potts_2019, Jorgensen_2020} [see Fig.~\ref{fig:coarsegraining}(a)]; alternatively, one addresses the sample indirectly, by measuring a probe that has interacted with it~\cite{Correa_2017, Hovhannisyan_2018, Miller_2018, Mitchison_2020, Henao_2020} [see Fig.~\ref{fig:coarsegraining}(b)].}
\mar{All of these cases are examples of a coarse-grained measurement, which from an abstract point of view can be described by a $d$-outcome \kar{generalized quantum} measurement of a $D$-dimensional system, with $d < D$. The fact that $d \ll D$ in most physically relevant cases may} reduce the precision significantly. It is hence natural to ask what the optimal measurement strategy and the associated precision of temperature estimation are under such limitations.
In this paper, we put forth a framework for addressing this question in detail. The framework is based on ideas from signal processing and parameter-estimation theory, and provides a simple, easy-to-use toolbox for studying coarse-grained thermometry of both few- and many-body systems.  We illustrate the framework by applying it to paradigmatic many-body models of spin lattices, both close to and far from criticality.

\mar{In a second part of the article, these abstract ideas are applied to probe-based temperature measurements. Here, the temperature of a sample is estimated by letting it interact with a probe (\kar{and possibly some auxiliary system}), and then measuring the \kar{probe}, as illustrated in Fig.~\ref{fig:coarsegraining}(b). These type of measurements are particularly appealing since they provide a natural way to overcome one of the main challenges in thermometry: the design of noninvasive measurements, 
, e.g., for ultra-cold atomic gases \cite{Spiegelhalder_2009, McKay_2010, Olf_2015, Hohmann_2016, Lous_2017, Bouton_2020}.
In such probe-based measurements, a natural strategy is to let the probe reach thermal equilibrium with the sample~\cite{Correa_2015}. Yet, it has been shown that the precision can be considerably enhanced by nonequilibrium strategies, where the probe either interacts with the sample for a finite time~\cite{Brunelli_2011, Brunelli_2012, Jevtic_2015, Guo_2015, Pasquale_2017, Cavina_2018, Mukherjee_2019, Mitchison_2020, Mancino_2020, Bouton_2020}, couples strongly to the sample \cite{Correa_2017, Hovhannisyan_2018}, or uses an ancillary catalytic system \cite{Henao_2020}. Here, we use our framework to obtain a fundamental bound on any such nonequilibrium strategy. \jon{We map the problem of probe-based thermometry to that of coarse-grained thermometry, and determine} the maximal sensitivity that can be obtained with a probe of dimension~$d$. We construct a specific fine-tuned sample-probe interaction that always saturates the bound, and notably show that it can also be reached in relevant experimental situations. In particular, when the sample and the probe are described by a harmonic oscillator and a qubit, respectively, the optimal nonequilibrium estimation scheme in the low-temperature regime can be obtained via the  Jaynes-Cummings Hamitonian. This is of direct relevance to temperature measurements of Bose-Einstein condensates \cite{Recati_2005} or micromechanical resonators \cite{Brunelli_2011, Brunelli_2012} via qubit probes.}


\section{Framework}
\label{sec:framework}

To be specific, we consider a system $S$ living in a $D$-dimensional Hilbert space, described by a Hamiltonian $H$, and initially in a canonical thermal state
\bea \label{thermal_state}
\tau = \frac{1}{Z} e^{-\beta H},
\eea
where $\beta = 1/T$ is the inverse temperature (we adopt units such that $k_{B}=1$) and $Z = \Tr e^{-\beta H}$ is the partition function. This is a family of states parameterised by the temperature, and the smallest variance in estimating this parameter, based on any measurement, is hence lower bounded by the quantum Cram\'{e}r-Rao bound \cite{Braunstein_1994} (see also \cite{Mandelbrot_1956, Uffink_1999})
\begin{equation} \label{cramer_rao}
    \Delta T^2 \geq \frac{1}{n \F(\tau)} ,
\end{equation}
where $n$ is the number of repetitions of the measurement 
and $\F$ is the quantum Fisher information with respect to $T$, which we refer to as the `thermal Fisher information'. It is given by
\begin{align} \label{thermal_fisher}
\F(\tau) = \beta^4 \var (H) = \beta^2 C,
\end{align}
with $\var (H) := \av{H^2}_\tau - \av{H}_\tau^2$, where the angle brackets denote averaging over the quantum state: $\av{\hat{O}}_\tau := \Tr (\hat{O} \tau)$, and $C$ is the heat capacity of the system ($C := d \av{H}/ d T$). The optimal measurement, attaining the thermal Fisher information, is a projective measurement onto the eigenbasis of $H$, i.e., a projective measurement of the system energy. In this optimal scenario, the more the energy fluctuates, the more precise the measurement can be in principle.  \jon{The optimal measurement saturates the inequality \eqref{cramer_rao} for arbitrary $n$ when the temperature estimator is unbiased.} \kar{When an unbisaed estimator is not available, a large class of generic biased estimators will still asymptotically saturate the inequality \eqref{cramer_rao} in the limit of many repetitions ($n \to \infty$) \cite{vanderVaart}. Moreover, in the specific case of temperature estimation in many-body systems consisting of $N \gg 1$ particles, one can go even further and explicitly construct an estimator that, despite being biased for any finite $N$, will saturate the Cram\'{e}r-Rao bound \eqref{cramer_rao} as $N \to \infty$, even for $n = 1$ \cite{Jahnke_2011}.}

\kar{However, when fine-grained measurements of energy are not available, while remaining valid, the bound \eqref{cramer_rao} will in general become too loose. We formalise the problem as follows. Suppose the resolution of the experiment is limited to $d < D$ measurement outcomes. \textit{What is then the maximal precision for estimating the temperature of $S$ and which is the optimal $d$-outcome measurement achieving it?}}
Below, we provide a systematic way to identify the optimal measurement.

\kar{Moreover, in Sec.~\ref{sec:noneqth}, we prove that the highest Fisher information achievable by a $d < D$ dimensional probe undergoing an arbitrary interaction with the sample is equal to the optimal $d$-outcome coarse-grained Fisher information in the above sense. We thus provide a fundamental benchmark for any conceivable protocol of probe-based thermometry. In particular, this includes any standard thermometric technique in current experimental setups.}

\section{Optimal coarse graining}
\label{sec:optimPOVM}

We consider coarse-grained thermometry on a $D$-dimensional system. We take coarse graining to mean that only generalised measurements, i.e., positive operator-valued measures (POVM) with at most $d < D$ outcomes are available. We then construct a framework for identifying the optimal POVM for thermometry in two steps.

First, we show that the optimal POVM is a projection onto energy subspaces of the system. This means that we can split the system spectrum into $d$ subsets and study measurements which project onto the corresponding eigensubspaces.

Second, we show that the optimal choice of subsets consists of consecutive ``bins,'' i.e., the sets are not interspersed. We then lay down a method for constructing the optimal choice of bins, for any given system spectrum. This is done by casting the problem in the language of an analogous signal-processing problem, known as Lloyd-Max quantization~\cite{Gallager}.

\subsection{Optimal POVM}
\label{sec:optipovm}

We take a $d$-outcome POVM $\M = \{ M_\alpha \}_{\alpha = 1}^d$ and a system in the thermal state $\tau$ of Eq.~\eqref{thermal_state}. Each outcome $\alpha$ then occurs with probability
\bea \label{palpha}
p_\alpha = \Tr(M_\alpha \tau) .
\eea
This distribution contains information about the temperature $T$, as quantified by the Fisher information \cite{Kullback}
\bea
\C = \sum_{\alpha = 1}^d \frac{1}{p_\alpha} \bigg( \frac{\partial p_\alpha}{\partial T} \bigg)^2,
\eea
which, for a thermal state, becomes \cite{Potts_2019}
\bea \label{eq:fiserinfo}
\C = \beta^4 \sum_\alpha \frac{1}{p_\alpha} \big[ \Tr \big(\tau M_\alpha (H - \av{H}_\tau) \big) \big]^2 .
\eea
\kar{Note that, after coarse graining, $p_\alpha$ is no longer in the so-called exponential family with respect to $T$\footnote{\kar{In that $p_\alpha$ cannot be written as $\exp[\mathfrak{K}(\alpha) f(T) + \mathfrak{L}(T) + \mathfrak{M}(\alpha)]$, where $f$, $\mathfrak{K}$, $\mathfrak{L}$, and $\mathfrak{M}$ are some functions.}}, which means that no temperature estimator can satisfy the Cram\'{e}r-Rao bound
for any finite number of repetitions $n$ \cite{vanderVaart}. However, the Fisher information ($\C$) is still a key precision quantifier as, for all unbiased and certain generic biased estimators, such as the maximum likelihood estimator, the Cram\'{e}r-Rao bound is saturated asymptotically in the $n \to \infty$ limit. Moreover, in the same limit, the Fisher information retains its key role even in the Bayesian estimation approach\footnote{\kar{Indeed, it follows from the Bayesian Cram\'{e}r-Rao bound (a.k.a. van Trees inequality) \cite{Gill_1995} that, for \textit{any} estimator, in the $n \to \infty$ limit, $\Delta T \gtrsim 1 / \big[n \int d T \, p(T) \C(T) \big]$, where $p(T)$ is the prior distribution of $T$.}}.}

\kar{We thus set as our goal} to determine the optimal~$d$-outcome POVM which maximises the Fisher information~$\C$.

We first observe that the POVM elements $M_\alpha$ can be taken to be diagonal in the energy eigenbasis. Because $\tau$ is diagonal in this basis, only diagonal elements of $M_\alpha$ contribute to the probability $p_\alpha$ in Eq.~\eqref{palpha}. Dropping all off-diagonal elements from each $M_\alpha$ results in a valid POVM since the operators remain positive and still sum to identity. Hence, for every POVM, there exists a diagonal POVM which achieves the same $p_\alpha$ and thus the same Fisher information. It is therefore sufficient to consider diagonal POVMs when looking for an optimal POVM maximising $\C$.

Next, let us note that $\C$ is convex with respect to the POVM. That is, denoting by $\C_\M$ the Fisher information corresponding to a particular POVM and considering two POVMs $\M$ and $\N$ and a mixing parameter $0\leq \lambda \leq 1$, we have
\bea \label{convexity}
\C_{\lambda \M + \overline{\lambda} \N} \leq \lambda \C_\M + \overline{\lambda} \C_\N,
\eea
where $\overline{\lambda} = 1 - \lambda$. This can be seen by rewriting Eq.~\eqref{eq:fiserinfo} as
\bea \label{partkom}
\C = \beta^4 \sum_\alpha \frac{1}{p_\alpha} \big(W_\alpha \big)^2 ,
\eea
where $W_\alpha = \Tr \big(\tau M_\alpha [H - \av{H}_\tau] \big)$. Both $p_\alpha$ and $W_\alpha$ are linear in the POVM operators. Hence, when mixing POVMs, $W_\alpha^{(\lambda \M + \overline{\lambda} \N)} = \lambda W_\alpha^{(\M)} + \overline{\lambda} W_\alpha^{(\N)}$ and $p_\alpha^{(\lambda \M + \overline{\lambda} \N)} = \lambda p_\alpha^{(\M)} + \overline{\lambda} p_\alpha^{(\N)}$. Equation \eqref{convexity} is then an immediate consequence of the fact that $\frac{1}{p} W^2$ is a jointly convex function of $p$ and $W$ (see, e.g., Ref.~\cite{Effros_2009}).

Finally, we show that the optimal POVM is necessarily a collection of nonoverlapping projectors onto eigensubspaces of $H$. Indeed, take a POVM $\M$ such that, for some eigenstate $\ket{k}$ of $H$, there are at least two POVM elements for which $\bra{k} M_{\alpha} \ket{k} > 0$. Define
\bea
\varsigma_\alpha = \bra{k} M_\alpha \ket{k} .
\eea
Now, construct $d$ new POVMs $\N^{(\gamma)}$, with elements
\bea
N^{(\gamma)}_\alpha = M_\alpha + (\delta_{\alpha \gamma} - \varsigma_\alpha) \ket{k}\bra{k}.
\eea
Each $\N^{(\gamma)}$ has the property that only $N^{(\gamma)}_\gamma$ has a nonzero $k$'th diagonal, namely $\bra{k} N^{(\gamma)}_\gamma \ket{k} = 1$ while $\bra{k} N^{(\gamma)}_{\alpha \neq \gamma} \ket{k} = 0$. Furthermore, we note that
\bea
\sum_\gamma \varsigma_\gamma \N^{(\gamma)} = \M .
\eea
Since the $M_\alpha$ form a POVM, we have $\varsigma_\alpha \geq 0$ and $\sum_\alpha \varsigma_\alpha = 1$. By convexity \eqref{convexity} of the Fisher information, 
\bea
\C_\M \leq \sum_\gamma \varsigma_\gamma \C_{\N^{(\gamma)}} ,
\eea
and it follows that there must be at least one $\gamma$ for which $\C_{\N^{(\gamma)}}(T) \geq \C_\M(T)$. This means that the optimal POVM will have to be one that consists of nonoverlapping projectors on eigensubspaces of $H$. Given a \gtl{(possibly degenerate)} Hamiltonian $H = \sum_i E_i \ket{i} \bra{i}$ the optimal POVM will thus be of the form
\begin{equation}\label{Pi_alpha}
    \Pi_\alpha = \sum_{E_i \in I_\alpha} \ket{i}\bra{i},
\end{equation}
where the $I_\alpha$ define a partition of the set of all eigenenergies into nonoverlapping subsets (``bins'').

To summarize, optimal, coarse-grained thermometry can always be achieved by considering projective measurements onto nonoverlapping eigenenergy subspaces. 

\subsection{Optimal binning}
\label{sec:optbinning}

We now construct a method for determining the optimal eigenenergy subsets, defining the optimal POVM.

\kar{For convenience, we choose the basis of $H$ such that
\bea \label{E_ordering}
E_1 \leq E_2 \leq \cdots \leq E_D,
\eea
and write the probability of finding the system in bin $I_\alpha$ as [cf. Eq.~\eqref{palpha}]
\bea \label{pk_def}
p_\alpha = \Tr(\Pi_\alpha \tau) = \sum_{E_i \in I_\alpha} q_i ,
\eea
where $q_{i} = \exp (-\beta E_{i})/Z$. Next, we introduce the ``bin energies'' (normalized average energy within each bin),
\bea \label{epsilon_k}
\epsilon_\alpha = \frac{1}{p_\alpha} \sum_{E_i \in I_\alpha} q_i E_i,
\eea
and, with these definitions, reexpress the Fisher information in Eq. \eqref{eq:fiserinfo} for the corresponding measurement as
\bea \label{eq:C_def}
\C = \beta^4 \sum_{\alpha = 1}^{d} p_\alpha (\epsilon_\alpha - \av{H})^2.
\eea
We shall henceforth refer to this as the \emph{coarse-grained} Fisher information.
Compared with Eq.~(\ref{thermal_fisher}), the expression for $\mathcal{C}$ describes how each of the energies $\epsilon_\alpha$ fluctuates away from the average.}

\kar{As a first step towards finding the optimal sets $I_\alpha$, in Appendix~\ref{app:binning}, applying a result from Ref.~\cite{Chakravarty_1985}, we prove that the best choice of $I_\alpha$ is given by a binning into consecutive intervals: $I_\alpha = \{E_{i_{\alpha - 1}}, E_{i_{\alpha - 1} + 1}, \ldots, E_{i_{\alpha - 1} + |I_\alpha| - 1}\}$, where $1 \leq i_{\alpha - 1} \leq D$ and the iteration rule is $i_\alpha = i_{\alpha - 1} + |I_\alpha|$. Introducing the ``boundaries'' $b_\alpha := E_{i_\alpha}$, we can conveniently write $I_\alpha = [b_{\alpha - 1}, b_\alpha)$, with the proviso that $b_0 = E_1$ and $b_d = E_{D + 1} = E_D$ (the extra level $E_{D + 1}$ will not enter into any physical quantity and is introduced just so that $E_D$ enters $I_d$ despite $[b_{d - 1}, b_d)$ having an open end). Note that, for discrete spectra, the boundaries $b_\alpha$ need not be exactly at energy eigenvalues. Positioning a boundary anywhere between neighboring eigenvalues results in the same POVM in Eq.~\eqref{Pi_alpha}.}

The remaining task is then to find the optimal intervals $I_\alpha = [b_{\alpha - 1}, b_\alpha)$ which maximize $\mathcal{C}$. This will give the best strategy for temperature estimation using a $d$-outcome POVM.

Before carrying out this optimization, it is  useful to recast the problem in terms of the density of states (DOS)
\begin{equation} \label{DOS}
    \Omega(E) = \sum\limits_i \delta(E- E_i),
\end{equation}
where $\delta$ denotes the Dirac's delta function (note that this definition does not assume a continuous spectrum). Expectation values of any function $g(H)$ of the Hamiltonian may then be written as 
\begin{equation}\label{eq:gH}
    \langle g(H) \rangle = \sum\limits_i g(E_i) q_i = \int dE g(E) \Omega(E) \frac{e^{-\beta E}}{Z}. 
\end{equation}
This means that we can define the distribution of energy as
\bea \label{energdist}
q(E) = \Omega(E) \frac{e^{-\beta E}}{Z},
\eea
so that expectation values can be computed simply in terms of integrals over $q(E)$. This way, the probabilities $p_\alpha$ in Eq.~\eqref{pk_def} and the bin energies in Eq.~\eqref{epsilon_k} can be conveniently written as 
\bea \nonumber
p_\alpha = \int_{b_{\alpha - 1}}^{b_\alpha} \! dE \, q(E) \quad \mathrm{and} \quad \epsilon_\alpha = \frac{1}{p_\alpha} \int_{b_{\alpha - 1}}^{b_\alpha} \! dE \, q(E) E. ~
\\ \label{pk_eps_int}
\eea
These quantities are therefore all expressed explicitly as functions of the boundaries $b_\alpha$.

The advantage of introducing the DOS is twofold. 
First, it allows for a unified treatment of Hamiltonians with discrete and quasi-continuous spectra (as one would expect in quantum many-body systems).
Second, it allows us to frame the problem in the language of signal processing. 
A common task in signal processing is to transmit a continuous function $q(E)$ through a channel. 
Often, in order to do so, the function must be discretized into a finite set of bins  $I_\alpha = [b_{\alpha - 1}, b_\alpha)$.
The question is then which choice of bins leads to the optimal transmission. This problem is solved by using the scheme known as Lloyd-Max quantization (see Ref.~\cite{Gallager}, Chapter 3). If one uses the mean squared variations of energy as a figure of merit, one sees that the maximization of the Fisher information [see Eq.~\eqref{eq:C_def}] becomes entirely analogous to this signal processing problem. 

To proceed, we introduce the quantity 
\begin{equation}
    \mathcal{D} = \beta^4  \sum_{\alpha = 1}^d\;\; \int\limits_{b_{\alpha - 1}}^{b_\alpha} dE q(E) (E - \epsilon_\alpha)^2.
\end{equation}
It can be directly verified that the thermal Fisher information given by Eq.~\eqref{thermal_fisher} can be decomposed as 
\begin{equation}
    \F = \mathcal{C} + \mathcal{D},
\end{equation}
which means that the task of maximizing $\mathcal{C}$, for a fixed $\F$, is tantamount to that of minimizing $\mathcal{D}$. 

The minimization can  be carried out in the usual way, by equating to zero the derivatives of $\mathcal{D}$ with respect to $b_\alpha$.
A straightforward calculation 
shows that the minima of $\mathcal{D}$ occur when the intervals $b_\alpha$ satisfy the implicit (and generally nonlinear) equations,\footnote{
\gtl{The solutions to this equation are also guaranteed to be an actual minimum of $\mathcal{D}$, never a maximum. First, $\partial^2 \mathcal{D}/\partial b_{\alpha} b_{\alpha'} =  0$ for $\alpha' \neq \alpha$. Second, at $b_\alpha = (\epsilon_{\alpha+1}+ \epsilon_\alpha)/2$, we have     $\partial^2 \mathcal{D}/\partial b_{\alpha}^2= 2  \beta^4(\epsilon_{\alpha+1} - \epsilon_\alpha) q(b_\alpha) \geqslant 0$, since $q(b_\alpha)\geqslant 0$ by construction and $\epsilon_{\alpha+1} \geqslant \epsilon_\alpha$ by hypothesis.
}}
\begin{equation}\label{implicit}
    b_\alpha = \frac{\epsilon_{\alpha + 1} + \epsilon_\alpha}{2}, \qquad \alpha = 1,\ldots, d-1. 
\end{equation}
These equations are implicit because $\epsilon_\alpha$ itself is a function of the $\{b_\alpha\}$ [Eq.~(\ref{epsilon_k})].
This summarizes the core of our framework.
Equation \eqref{implicit} provides a recipe for how to optimize the energy bins in a $d$-outcome POVM in order to maximize the thermometric precision.

\subsection{Illustrative examples}
\label{sec:examples}

Let us now present two examples using our framework for optimal coarse-grained thermometry. 

\subsubsection{Noninteracting qubits}
\label{sec:Ntlss}

\begin{figure}
    \begin{center}
	\includegraphics[width=8cm]{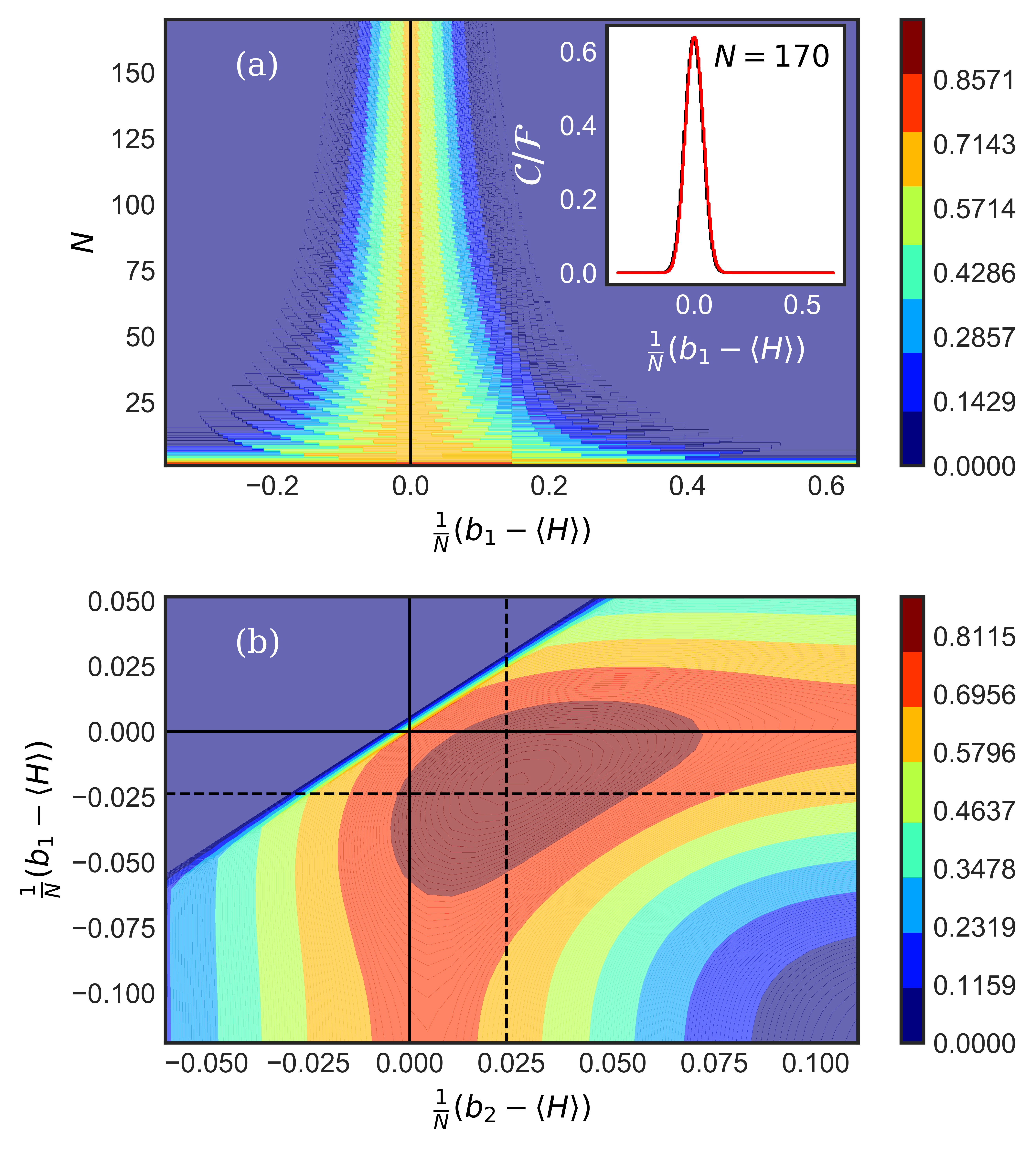}
	\caption{Ratio between the coarse-grained and thermal Fisher informations $\mathcal{C}/\F$ at $\beta=0.6$.
	(a) The case of $d=2$, plotted as a function of the partition $(b_1 - \av{H})/N$ and the number of qubits $N$. The ratio reaches a maximum value of approximately $0.64$ in the large $N$ limit.
	(Inset) Cross section at $N=170$.
	(b) The case of $d=3$ for $N=170$, plotted as a function of the lower partition $(b_{1} - \av{H})/N$ and the upper partition $(b_{2} - \av{H})/N$.
	The ratio has a maximum value of approximately $0.82$.
	The dashed lines gives the optimal partitions as predicted by a symmetric partition around the mean energy.}
    \label{fig:N-qubit}
    \end{center}
\end{figure}

A simple, but illustrative example, is a system of $N$ identical, noninteracting qubits. The system is in a thermal state, and take the ground- and excited-state energies to be 0 and 1, respectively. The energy levels of the system will thus range from $0$ to $N$ in integer steps. 
The probability to find the system in a state with energy $j$ is then
\begin{equation} \label{eq:NqubitEdist}
    q_j = {N \choose j} s^j r^{N-j},
\end{equation}
where we have defined the excited-state population $s = 1/(e^{\beta}+1)$ and $r=1-s$. For a $d$-outcome measurement, the probabilities and bin energies take the form
\beaa
\label{N_qubits_p_epsilon}
    p_\alpha &= \sum_{j = b_{\alpha - 1}}^{b_\alpha - 1} {N \choose j} s^j r^{N-j},
    \\
    \epsilon_\alpha &= \frac{1}{p_{\alpha}} \sum_{j = b_{\alpha - 1}}^{b_\alpha - 1} {N \choose j} j s^j r^{N-j},
\eeaa
where the bin positions $b_\alpha$, which are integers in this case, are determined from Eq.~\eqref{implicit} (with $b_{0} = 0$). 
Note that, for this system, the average energy is simply $\av{H} = N s$, while the thermal Fisher information~\eqref{thermal_fisher} is  $\F = \beta^{4} N r s$. In Fig.~\ref{fig:N-qubit}, we show the ratio between $\mathcal{C}/\mathcal{F}$ as a function of the bin positions $b_\alpha$, for the cases $d = 2, 3$. 
The bins $b_\alpha$ which maximize $\mathcal{C}/\mathcal{F}$ are precisely the solutions of
Eq.~\eqref{implicit}. 

According to the De Moivre--Laplace theorem, in the $N \gg 1$ limit, the distribution $q_j$ becomes Gaussian. Using this, we show in Appendix \ref{app:Nqubits} that, for binary measurements ($d = 2$) and $N \gg 1$, optimal binning strategy leads to 
\bea \label{NqubCF}
\C = \frac{2}{\pi}\F,
 \eea
and is achieved when the boundary is inserted at $b = \av{H} = N s$.
This result is noteworthy, it shows that, irrespective of the number of qubits $N$, it is always possible to construct a dichotomic measurement strategy which captures $(2/\pi) \approx 0.63$ of the full thermal Fisher information.

\subsubsection{\label{sec:linear_DOS}Linear density of states}

To contrast with the $N$ qubits case, we now consider an example of a system with a continuous spectrum. Namely, we assume the system has a linear density of states: $\Omega(E) \propto E$. Such a DOS is met, for instance, in a noninteracting, ultra-relativistic gas in two dimensions. If the system is in a thermal state [Eq.~\eqref{thermal_state}], the average energy is simply $\av{H} = 2/\beta$ and the variance is $\var(H) = 2/\beta^2$. Thus, the thermal Fisher information given in Eq.~\eqref{thermal_fisher} is $\F = 2\beta^2$. 

\begin{figure}
    \centering    
    \includegraphics[width=0.4\textwidth]{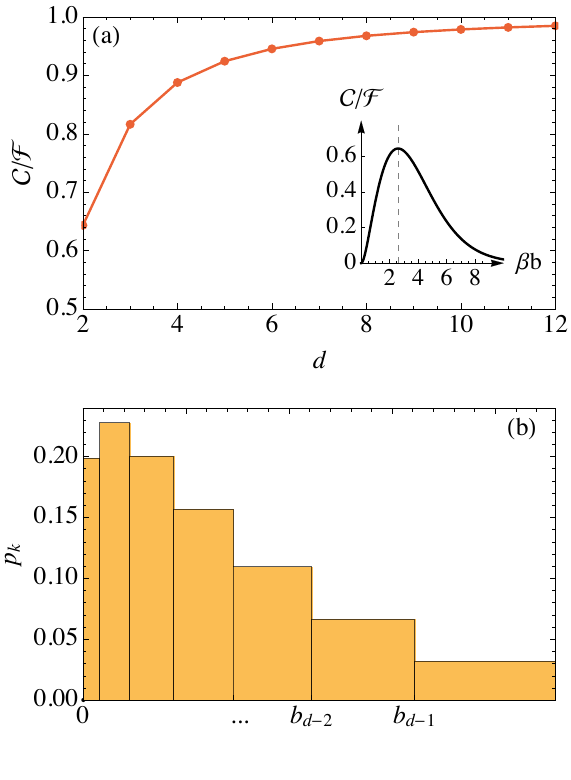}
    \caption{Optimal thermometry binning for a system described by a linear density of states: $\Omega(E) \propto E$. 
    (a) Optimal binned Fisher information, Eq.~(\ref{eq:C_def}), as a function of $d$. The curves are normalized by the full Fisher information, which in this case reads $\F = 2\beta^2$. 
    Inset: ratio $\mathcal{C}/\F$ as a function of the binning position $b$ for $d = 2$. The optimal bin occurs at $\beta b = 2.58975$. The points in the main plot are already optimized over the binning positions. 
    (b) Illustration of the corresponding probabilities $p_k$ and average bins $b_k$ for $d = 8$.}
    \label{fig:linearDOS}
\end{figure}

We first consider the case of binary measurements, $d = 2$, defined by a single boundary $b$. The probabilities $p_1$ and $p_2$ [Eq.~\eqref{pk_eps_int}] are then given by
\bea \nonumber  
p_1 = 1 - p_2, \qquad p_2 = \int\limits_{b}^\infty dE \, q(E) = (1 + \beta b) e^{-\beta b} , 
\eea
and the corresponding bin energies become
\bea \nonumber
    \epsilon_1 = \frac{2}{\beta} + \frac{\beta b^2}{1+ b \beta -e^{b \beta}}, \qquad
    \epsilon_2 = \frac{2}{\beta} + \frac{\beta b^2}{1+\beta b}.
\eea
Thus, the Fisher information for the measurement, Eq.~\eqref{eq:C_def}, is
\begin{equation}\label{C_linearDOS}
    \mathcal{C} = \frac{\beta^6 b^4}{(1+ \beta b)(e^{\beta b} -1 - \beta b)} .
\end{equation}
To find the optimal partition, we solve Eq.~\eqref{implicit} for $b$, i.e., $b = (\epsilon_1 + \epsilon_2)/2$. This is a nonlinear equation which can be solved numerically. A plot of 
$\mathcal{C}/\F$ is shown in the inset of Fig.~\ref{fig:linearDOS}(a). It attains a maximum at $\beta b \approx 2.589$. At this point $\mathcal{C} \approx 0.643 \F$, i.e., the binary measurement reaches approximately 64\% of maximal Fisher information for any possible measurement (this is slightly higher than in Eq.~\eqref{NqubCF}).

The dependence of $\mathcal{C}$ on the number of outcomes $d$, for optimal binnings, is shown in Fig.~\ref{fig:linearDOS}(a). 
Quite remarkably, even with as little as 8 bins, one can already reach a precision of $\approx 97\%$ of $\F$---the maximal possible precision. 
An illustration of the probabilities $p_k$ and the corresponding bins $b_k$ is given in Fig.~\ref{fig:linearDOS}(b) for $d = 8$.

\subsection{General remarks and extension to imperfect measurements}
\label{sec:Gremarks}

The two examples in Sec.~\ref{sec:examples} carry an important general message: even measurements so coarse-grained as to have only two outcomes can estimate the temperature of a generic system with precision (as quantified by the Fisher information) that is proportional to the ultimate precision---the thermal Fisher information [Eq.~\eqref{thermal_fisher}]. In Appendix \ref{app:propN}, we show that this is not a coincidence, by proving that any system for which the energy distribution is unimodal and has sufficiently fast decaying tails, displays a proportionality $\mathcal{C} \propto \mathcal{F}$. More specifically, we prove that there exists a finite number $\Xi \in [0,1]$ such that $\C \geq \Xi \F$.

As will be discussed in Sec.~\ref{sec:manybody}, the relevance of these results lies in the fact that unimodal energy distributions with quickly decaying tails are a generic behaviour expected in finite-temperature many-body systems with short-range interactions, both at and away from criticality. In fact, we will show that  $\Xi = 2/\pi$, as in Eq.~\eqref{NqubCF}, actually happens for a large variety of interacting lattice models. 
It is of course possible to conceive nonunimodal energy distributions for which the proportionality $\C \propto \F$ breaks down. We illustrate this in Appendix \ref{app:smallC}, where we construct an example for which $\mathcal{C}/\mathcal{F}\to 0$ when $N \to \infty$.

{\color{black}
\jon{Our framework can also be adapted to scenarios where the ideal energy binning cannot be implemented and imprecisions are present.}
In the simplest case, one could have some imprecision in determining the optimal bins through Eq.~\eqref{implicit}\jon{, which would lead to a decrease in $\mathcal{C}$. For the case $d=2$, this is illustrated in the inset of Fig.~\ref{fig:linearDOS}(a), which shows how $\mathcal{C}/\mathcal{F}$ decreases as the bin position deviates from its optimal value.} 

Another way in which imprecisions can enter our framework is when the POVMs~\eqref{palpha} are noisy. For instance, experimental errors may cause energies close to a bin edge to sometimes result in outcomes corresponding to neighbouring bins.
Such effects can be accounted for within our framework by modifying Eq.~\eqref{pk_eps_int}. 
To see that, we first rewrite 
Eq.~\eqref{pk_eps_int} as
\bea \label{pk_eps_int_noisy}
p_\alpha = \int_{-\infty}^{\infty} \! dE \, q(E) B_\alpha(E), 
\eea
and similarly for $\epsilon_\alpha$. 
Here $B_\alpha(E)$ is a boxcar function, with value 1 when $E \in I_\alpha = [b_{\alpha-1}, b_\alpha)$, and 0 otherwise. 
A similar analysis can also be done at the discrete representation of Eq.~\eqref{pk_def}. In this case, we would have $p_\alpha = \sum_i B_{\alpha,i} q_i$, where $B_{\alpha,i}$ is a matrix with entries 1 when $E_i \in I_\alpha$ and zero otherwise. 
However, it is more convenient to work with the continuous-energy representation~\eqref{pk_eps_int_noisy}. 

It is now straightforward to generalize Eq.~\eqref{pk_eps_int_noisy} to include the effects of noise by replacing $B_\alpha(E)$ by a different function. For instance, a smoothed boxcar as depicted in Fig.~\ref{fig:MessyBins}(a). 
Since $\sum_\alpha p_\alpha = 1$ for any initial distribution $q(E)$, it follows that $\sum_\alpha B_\alpha(E) = 1$ for all $E$.
This can be viewed as a normalization condition for $B_\alpha(E)$. In fact, $B_\alpha(E)$ is actually a combination of a stochastic matrix (whose columns add up to one), plus an isometry, which reduces the dimension from a continuous energy $E$, to a discrete set of outcomes $\alpha$. The precise form of $B_\alpha(E)$ will depend on the details of the experiment.

Measurement errors can cause not only a loss of precision, but also systematic shifts in energy by, e.g., falsely displacing the energies $\epsilon_\alpha$ by a certain amount. For simplicity, we shall \kar{study these kinds of imprecision separately. We defer the discussion of robustness to energy shifts to Sections \ref{sec:gaussianEdist} and \ref{sec:noncrit}, while here we choose} $B_\alpha(E)$ to be symmetric in the interval $[b_{\alpha-1}, b_\alpha)$, and centered at the midpoint $(b_{\alpha-1} + b_\alpha)/2$\kar{, so that the $\epsilon_\alpha$'s are not displaced}.

The remaining feature to describe is errors associated with imperfect binning. This can again be done using the smoothed boxcar of Fig.~\ref{fig:MessyBins}(a)
\begin{equation}\label{smoothed_boxcar}
    B_\alpha(E) = \frac{1}{2} \erf \left(\frac{b_\alpha - E}{\sigma\sqrt{2}}\right) - \frac{1}{2} \erf \left(\frac{b_{\alpha-1} - E}{\sigma\sqrt{2}}\right),
\end{equation}
where $\erf(x)$ is the error function and $\sigma$ is a parameter measuring the degree of imprecision (a sharp boxcar is recovered when $\sigma \to 0$). 
A function of this form defines a certain energy window $2\sigma$, where measurements associated to one bin can be recorded in another. 
For this reason, the wider bins tend to be less affected than the thinner ones, which is physically reasonable. 

\begin{figure}
    \centering
    \includegraphics[width=0.45\textwidth]{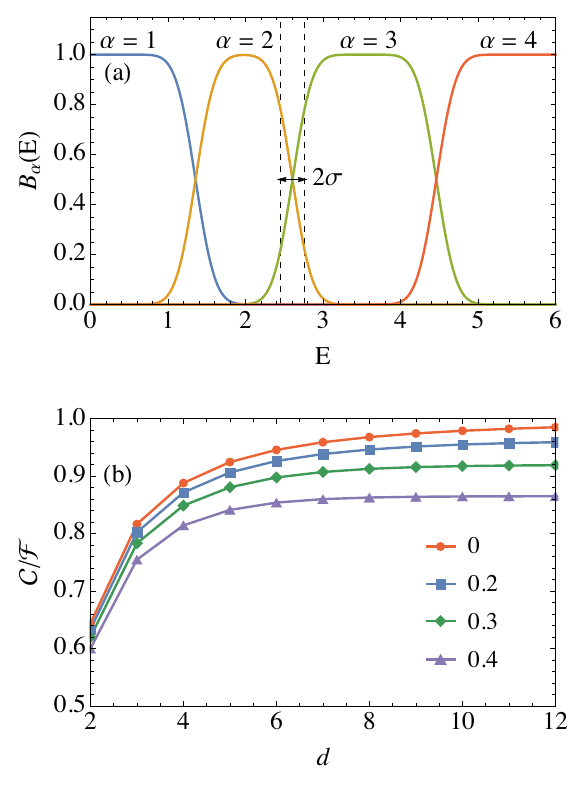}
    \caption{\color{black}
    The role of imperfect measurements in optimal thermometry.
    (a) Example if smoothed boxcars, Eq.~\eqref{smoothed_boxcar}, for $\sigma = 0.2$.
    (b) Illustration of the worsening of the coarse-grained Fisher information $\mathcal{C}$ for the linear DOS example studied in Sec.~\ref{sec:linear_DOS}. The red curve is the same as in Fig.~\ref{fig:linearDOS}(a), while the other curves were computed using the smoothed boxcars~\eqref{smoothed_boxcar} with $\sigma = 0.2$, $0.3$ and $0.4$ (in units of $\beta = 1$).
    }
    \label{fig:MessyBins}
\end{figure}

We illustrate the above ideas with the linear DOS example of Sec.~\ref{sec:linear_DOS}.
In Fig.~\ref{fig:MessyBins}(b) we present $\mathcal{C}/\mathcal{F}$ as a function of $d$ for binning strategies computed using the smoothed boxcars~\eqref{smoothed_boxcar}, with different values of $\sigma$. 
This is contrasted with the ideal case, shown in red, which coincides with the red curve in Fig.~\ref{fig:linearDOS}(a).
As can be seen, unsharp bin edges necessarily decrease the coarse-grained Fisher information. 
That said, $\C/\F$ is surprisingly robust: even when the smearing occurs over a large part of the bin width \kar{(e.g., $30\%$)}, $\mathcal{C}/\mathcal{F}$ does not decrease much \kar{(only about $7\%$)}.


}


\section{Many-body lattice models}
\label{sec:manybody}

We now proceed to analyze quantum systems on a lattice, which is one of the most physically relevant settings where the coarse-grained measurements could be useful. We start with general considerations and a tight-binding chain as an illustrative example. Then we show a general result for all noncritical spin models and conclude with an analysis of a system undergoing a thermal phase transition.

\subsection{Gaussian density of states}
\label{sec:gaussianEdist}

In many-body lattice models, the energy distribution~\eqref{energdist} often displays an approximate Gaussian form in the thermodynamic limit \cite{Keating_2015, Brandao_2015, Brandao_2015G} (see also the detailed discussion in Sec.~\ref{sec:noncrit} and Appendix \ref{app:many}). As a simple, illustrative example of this principle,  consider a fermionic one-dimensional tight-binding chain with $N$ sites, under periodic boundary conditions:
\begin{equation}
    H = \sum_{k = 1}^N \varepsilon \hat{c}_k^\dagger \hat{c}_k - t \sum_{k = 1}^N (\hat{c}_{k + 1}^\dagger \hat{c}_k + \hat{c}_k^\dagger \hat{c}_{k + 1}),
\end{equation}
where $\hat{c}_k$ is the fermionic annihilation operator at site $k$, $\varepsilon$ is the on-site energy, $t$ is hopping (tunneling) strength, and $\hat{c}_{k + N} = \hat{c}_k$ ensures periodic boundary conditions. When diagonalized, the Hamiltonian of this model takes the form \cite{Simon, Potts_2019}
\begin{equation}
    H = \sum_{a = 1}^{N} \varepsilon_{a} \hat{C}_{a}^{\dagger} \hat{C}_{a} ,
\end{equation}
with the (linearly) transformed $\hat{C}_a$'s satisfying standard fermionic anti-commutation relations, and with eigenenergies given by $\varepsilon_{a} = \varepsilon - t \cos (2 \pi a / N)$.

\begin{figure}
    \centering
    \includegraphics[width=0.45\textwidth]{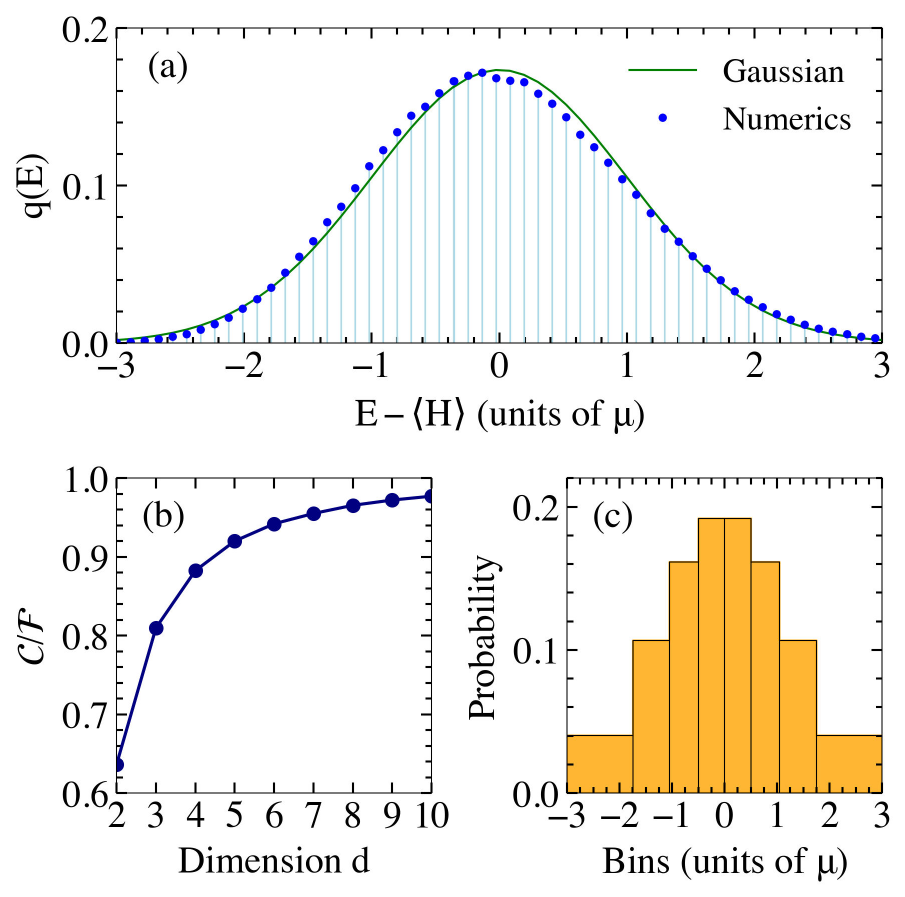}
    \caption{
    (a) Energy distribution for a fermionic tight-binding chain with 20 sites. The hopping strength is $t=0.3$, and the temperature is $T/\epsilon = 2$.
        The numerical results are compared to the Gaussian distribution (Eq.~\eqref{eq:gaussian}).
    (b) Optimal binned Fisher information for the general Gaussian distribution as a function of $d$. The curves are normalized by the thermal Fisher information (Eq.~\eqref{thermal_fisher}).
    (c) Illustration of optimal binning, and the corresponding probabilities, for the general Gaussian distribution with $d=8$.}
    \label{fig:gaussianDOS}
\end{figure}

In Fig.~\ref{fig:gaussianDOS}(a), we numerically compute the energy distribution~\eqref{energdist} of this model, and compare the results to a continuous Gaussian distribution with average energy $\av{H}$ and variance $\mu^{2} \equiv \var(H) = -\partial\av{H}/\partial\beta$ (both of which depend implicitly on $T$); i.e.,
\begin{equation}\label{eq:gaussian}
    q(E) = \frac{1}{\sqrt{2 \pi \mu^{2}}} e^{-\frac{(E - \av{H})^2}{2 \mu^{2}}}.
\end{equation}
We observe that the Gaussian distribution is a good approximation already with a modest number of sites and a modest hopping strength. 
The approximation, in fact, improves with the number of sites and becomes exact  in the thermodynamic limit (see Refs.~\cite{Brandao_2015, Brandao_2015G}).

Let us now take a Gaussian distribution as a given and compute the  Fisher information for different coarse-grainings of the continuous distribution \eqref{eq:gaussian}.
In this case, the probabilities and bin energies in~\eqref{pk_eps_int} become
\beaa
    p_{\alpha} &= \frac{1}{2} \big[ \erf \! \big( \, \tilde{b}_{\alpha} \big) - \erf \! \big( \, \tilde{b}_{\alpha-1} \big) \big] ,
\\
    \epsilon_{\alpha} &= \av{H}
    - \frac{\mu}{p_{\alpha} \sqrt{2 \pi}} \Big( e^{-\tilde{b}_{\alpha}^{2}} - e^{-\tilde{b}_{\alpha-1}^{2}} \Big) ,
\eeaa
where $\tilde{b}_{\alpha}:= (b_{\alpha} - \av{H}) / \sqrt{2} \mu$ (with $b_0 = -\infty$, $b_d = \infty$) are the shifted and rescaled bin positions. The full Fisher information is simply $\mathcal{F} = \beta^4 \mu^2$. For a $d$-outcome measurement, one can then numerically perform the optimization according to Eq.~\eqref{implicit} to find the best such measurement and the corresponding coarse-grained Fisher information~\eqref{eq:C_def}. The results for $\mathcal{C}/\mathcal{F}$  for different numbers of bins $d$, are shown in Fig.~\ref{fig:gaussianDOS}(b).
As in the linear density of states case, one sees a quick growth of $\mathcal{C}$ with $d$ towards the maximum value $\mathcal{F}$. 

The particular case of $d = 2$ can be obtained by setting $b_0 = -\infty$, $b_2 = \infty$ and $b_1 = b$. We then get 
\begin{equation} \nonumber
p_{1, 2} = \frac{1 \pm \erf(\tilde{b})}{2}, \qquad \epsilon_{1,2} = \av{H} \mp \frac{\mu}{\sqrt{2 \pi} p_{1, 2}} e^{-\tilde{b}^2},
\end{equation}
so that the coarse-grained Fisher information [Eq.~\eqref{eq:C_def}]
becomes
\begin{equation} \label{C_gaussian_b}
\mathcal{C} = \frac{2 \mathcal{F}}{\pi} \frac{e^{-2 \tilde{b}^2}}{1 - [\erf (\tilde{b})]^2}.
\end{equation}
The result is expressed solely in terms of the shifted bin position $\tilde{b}$; therefore, the minimization procedure is independent of $\av{H}$ or $\mu$. In fact, as one may readily verify, the function in Eq.~\eqref{C_gaussian_b} is maximized at $\tilde{b} = 0$. That is, the bin should be placed symmetrically, at $b = \av{H}$. 
The corresponding maximum is 
\begin{align} \label{qfidim2}
\mathcal{C} = \frac{2}{\pi} \F .
\end{align}
\kar{This relation is robust with respect to imprecise identification of the optimal boundary (which can be understood as a systematic error in the energy measurement, as mentioned in Sec.~\ref{sec:Gremarks}). Indeed, Taylor-expanding the right-hand side of Eq.~\eqref{C_gaussian_b} with respect to small $\tilde{b}$ around it optimal value, $\tilde{b} = 0$, we find that $\frac{\C}{\F} = \frac{2}{\pi} [1 - 2 \tilde{b}^2 (1 - 2/\pi) + O(\tilde{b}^4)]$. Even for a significant deviation of $|b - \av{H}| = 0.3 \mu$, $\C / \F$ degrades only by $\approx 3.3 \%$.}

\kar{Not coincidentally, the relation in Eq.~\eqref{qfidim2} also} appears in the case of noninteracting qubits in the limit of large $N$ (Sec.~\ref{sec:Ntlss}). 
This is because the energy distribution in that case also becomes Gaussian in the $N \gg 1$ limit, due to the central limit theorem.

In Fig.~\ref{fig:gaussianDOS}(c), we illustrate the optimal bins and the corresponding probabilities for the distribution \eqref{eq:gaussian} in the case of $d = 8$. In this case, the optimal bins have to be located numerically. Unsurprisingly, it is found that the optimum is symmetric around the average energy.

\subsection{Noncritical, interacting systems on lattices}
\label{sec:noncrit}

Now we will show how some of the conclusions of Sec.~\ref{sec:gaussianEdist} actually hold universally in the thermodynamic limit. 
Intuitively speaking, the idea is that generic lattice models with finite-range interactions, when away from criticality, tend to have Gaussian energy distribution due to the many-body Berry-Esseen theorem \cite{Keating_2015, Brandao_2015, Brandao_2015G}. Therefore, the same behaviour as in Fig.~\ref{fig:gaussianDOS} is expected to occur when coarse-graining to different partitions in such lattice models. 

\kar{In fact, in Appendix \ref{app:many}, we prove that the maximal $\C / \F$ achievable by two-outcome measurements ($d = 2$) is $\frac{2}{\pi} + O(\ln^{-1} N)$, and the boundary of the optimal partition $I_1 = [E_0, b]$ and $I_2 = (b, E_D]$ is near the average energy: $b - \av{H} = O(\ln^{-1/2} N) \sqrt{\var(H)}$; moreover, when $b = \av{H}$ exactly, one still has $\frac{\C}{\F} = \frac{2}{\pi} + O(\ln^{-1} N)$.}
\kar{In the the thermodynamic limit,} this coincides with the results for the exact Gaussian distribution (Sec.~\ref{sec:gaussianEdist}) and for independent qubits (Sec.~\ref{sec:Ntlss}). We expect that, for $d \geq 3$ partitions, one should be able to prove results identical to those obtained for the exact Gaussian energy distribution in Sec.~\ref{sec:gaussianEdist} by using arguments along the lines of those in Appendix~\ref{app:many}.

In order to prove Eq.~\eqref{qfidim2}, we need to assume that the thermal state of the lattice satisfies the following two generic conditions:

(i) Exponential decay of correlations: For arbitrary regions $\mathcal{X},\mathcal{Y}$ separated by a distance $l$ on the lattice, and some constant $\xi$,
\begin{equation}
    \max_{X\in \mathcal{X},Y\in \mathcal{Y}} \left | \frac{\langle X \otimes Y \rangle - \langle X \rangle \langle Y \rangle}{\vert \vert X \vert \vert \, \vert \vert Y \vert \vert} \right | \le e^{-l/\xi}.
\end{equation}

(ii) The variance in energy scales linearly with the number of lattice sites: $\var(H) = \av{H^2} - \av{H}^2 = s^2 N$.

Assumption (i) is expected to hold for a very large class of systems away from criticality. Indeed, it has been rigorously proven for $1$D translation-invariant thermal states \cite{Araki_1974}, finite-range fermionic lattice systems of arbitrary spatial dimension at nonzero temperatures \cite{Hastings_2004F}, and all finite-range lattice systems above a threshold lattice-dependent temperature \cite{Kliesch_2014}.
Assumption (ii) is expected to hold for most systems at a high enough temperature. In fact, note that (i) already implies that $\var(H) = O(N)$ \cite{Farrelly_2017}.

The detailed proof of Eq.~\eqref{qfidim2} in Appendix~\ref{app:many} is based on the Berry-Esseen theorem for local Hamiltonians which relies on the two assumptions above and is proven in Refs.~\cite{Brandao_2015, Brandao_2015G} (see also Appendix \ref{app:away}). This result can be seen as a strengthening of the central limit theorem, which gives a precise notion of how the energy distribution of lattice models converges to a Gaussian in the thermodynamic limit. It allows us to estimate functions of the form of Eq.~\eqref{eq:gH}, in this case, the bin energies $\epsilon_k$.

\kar{Lastly, the Gaussian behavior of noncritical many-body lattice systems also extends to the problem of how robust $\C / \F = 2 / \pi$ is to imprecise identification of the optimal binning boundary. Indeed, as we show in Appendix~\ref{app:many}, for small $\tilde{b} = \frac{b - \av{H}}{\sqrt{2 \var(H)}}$,
\bea \nonumber
\frac{\C}{\F} = \frac{2}{\pi} [1 - 2 \tilde{b}^2 (1 - 2 / \pi) + O(\ln^{-1} N) + O(\tilde{b}^3)],
\eea
so, as in Sec.~\ref{sec:gaussianEdist}, for e.g. $|b - \av{H}| = 0.3 \sqrt{\var(H)}$ (and $N \gg 1$), $\C / \F$ will degrade only by $\approx 3.3 \%$.}

\subsection{Critical systems}
\label{sec:crit}

The thermal Fisher information~\eqref{thermal_fisher} is proportional to the heat capacity of the system, which scales as $C = N c(\beta)$, where $c(\beta)$ is the specific heat. For noncritical systems, $c(\beta)$ is intensive. However,  
at a finite-temperature phase transition, it diverges as the temperature of the system approaches the critical temperature $T_c > 0$, according to $c(\beta) \propto |t|^{-\alpha}$, where $t := (\beta - \beta_c) / \beta_c$ and $\alpha \geq 0$ is the so-called critical exponent \cite{Fisher_1967}. When $\alpha = 0$, the divergence is logarithmic: $c(\beta) \propto \ln |t|^{-1}$ \cite{Fisher_1967}. In large but finite systems, there are of course no infinities and, at the phase transition point, $c_N(\beta_c) \propto N^{\frac{\alpha}{2 - \alpha}}$ when $\alpha > 0$ \cite{Brezin_1982} and $c_N(\beta_c) \propto \ln N$ when $\alpha = 0$ \cite{Brezin_1982, Izmailian_2002}.
Since $c_N = \beta^2 \var(H) / N$ [Eq.~\eqref{thermal_fisher}], the divergence of $c_N(\beta_c)$ with $N$ implies that critical systems do not satisfy the condition (ii) of Sec.~\ref{sec:noncrit}. In general, critical systems also feature diverging correlation lengths \cite{Fisher_1967}, thereby violating the condition (i) of Sec.~\ref{sec:noncrit} as well. Therefore, the many-body Berry-Esseen theorem becomes inapplicable for critical systems.

In Appendix \ref{app:critdistr}, building on several rigorous results on translation-invariant lattices with finite-range interactions in Refs.~\cite{Griffiths_1964, Simon, Muller_2015}, we develop an alternative approach
towards determining the energy distribution of such lattices in arbitrary spatial dimensions. Fist of all, for noncritical lattices, we show that the energy distribution approximates a Gaussian in a way that complements the many-body Berry-Esseen theorem \cite{Brandao_2015, Brandao_2015G}. Moreover, for this wide but specific class of lattices, our approach allows us to access the energy distribution even at criticality.

For critical lattices with $\alpha = 0$, we show in Appendix \ref{app:alphazero} that the energy distribution still tends to a Gaussian in the $N \to \infty$ limit; however, the convergence does not include the tails of the distribution, which are $O(\sqrt{N})$ standard deviations away from $\av{H}$. 
In a sense, for translation-invariant lattices, this result suggests that the Gaussianity of the distribution holds beyond assumptions (i) and (ii) above \cite{Brandao_2015, Brandao_2015G}.
Thus, Eq.~\eqref{qfidim2} applies in the thermodynamic limit, both at criticality (with $\alpha = 0$) and away from it. 
We illustrate these ideas below in Sec.~\ref{sssec:2dising}, with a detailed study on the classical 2D Ising model on a square-lattice, a paradigmatic model with $\alpha = 0$.

The case $1 > \alpha > 0$ is treated in Appendix \ref{app:positivealpha}. We show that the energy distribution is Gaussian only in a neighbourhood of the peak that is much smaller than the standard deviation. Hence, it is not Gaussian as a whole. 
Notwithstanding, we show that it is unimodal with exponentially decaying tails, which means that the considerations in Appendix \ref{app:propN} are applicable; that is, a two-bin measurements with the boundary placed at $\av{H}$ will yield a $\C$ that scales proportionally to $\F$. In other words, since $\F = \beta^2 N c_N$, we will have $\C \propto \beta^2 N^{\frac{2}{2 - \alpha}}$.

\subsubsection*{2D Ising model}
\label{sssec:2dising}

\begin{figure*}[!t]
\centering
\includegraphics[width=\textwidth]{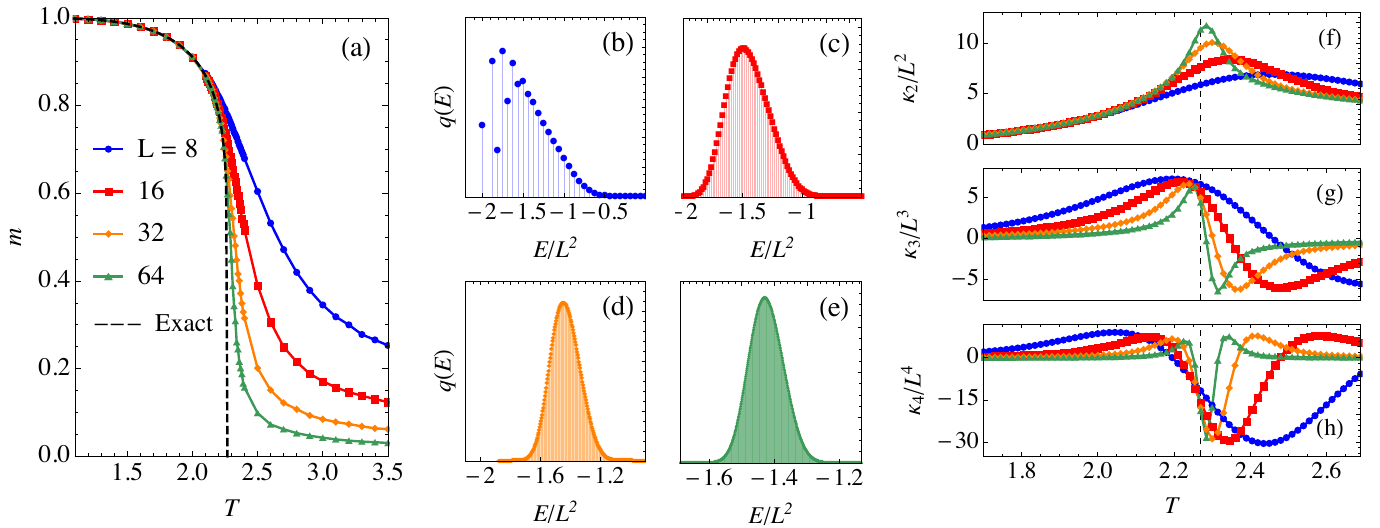}
\caption{Classical two-dimensional Ising model [Eq.~\eqref{Ising_energies}].
(a) Magnetization as a function of temperature for different lattice sizes, showing the phase transition at $T_c/J = 2.26919$. The dashed curve corresponds to the thermodynamic limit (where the exact solution is $m = \big[1 - \sinh^{-4}(2J/T)\big]^{1/8}$ \cite{Schultz_1964}).
(b)-(e) Exact energy distributions $q(E)$ for $T = T_c$, computed using the method of Ref.~\cite{Beale_1996}. Each curve corresponds to a different lattice size, $L = 8$, $16$, $32$, and $64$ (same color code as image (a)). 
(f)-(h) Cumulants $\kappa_2, \kappa_3$, and $\kappa_4$ of $q(E)$, as a function of temperature, for different lattice sizes.
The the third and fourth cumulants scale, respectively, as $N^{3/2}$ and $N^2$, where $N = L^2$ is the number of sites. The second cumulant, on the other hand, scales as $N\ln N$.
}
\label{fig:IsingGrid}
\end{figure*}

The square-lattice $2$D Ising model is defined on an $L\times L$ square lattice where each site $i$ is characterized by a Pauli matrix $\sigma_z^i$, with $i = 1, \ldots, N$ ($N = L^2$).
The spins interact according to the Hamiltonian 
\begin{equation}
    H = - J \sum\limits_{\langle i, j \rangle} \sigma_z^i \sigma_z^j,
\end{equation}
where the sum is over all nearest neighbors. Since the interactions only involve $\sigma_z$ operators, the Hamiltonian is already diagonal in the computational basis, with energy eigenvalues
\begin{equation}\label{Ising_energies}
    E(\bm{\sigma}) = - J \sum\limits_{\langle i, j \rangle} \sigma_i \sigma_j, 
\end{equation}
where $\bm{\sigma} = (\sigma_1, \ldots, \sigma_N)$ and $\sigma_i = \pm 1$ are the eigenvalues of $\sigma_z^i$. Here, we will impose periodic boundary conditions.
The model presents a phase transition at $T_c/J = 2/\ln(1+\sqrt{2})$ \cite{Kramers_1941, Onsager_1944, Schultz_1964}. This can be seen, for instance, in terms of the magnetization $m = \frac{1}{N}\sum_i \langle \sigma_i \rangle$, as plotted in Fig.~\ref{fig:IsingGrid}(a).

For not very large $N$'s, the full energy distribution $q(E)$ can be computed exactly using a method developed in Ref.~\cite{Beale_1996}. Results for $L = 8, 16, 32,$ and $64$ are shown in Fig.~\ref{fig:IsingGrid}(b)-(e). Although irregular for small sizes, it can be seen that the distribution visually appears to approach a Gaussian as the lattice size increases.

In order to rigorously prove that this is indeed the case, one needs to show that the higher-order ($\geq 3$) cumulants of the energy distribution, $\kappa_k$, become irrelevant as $N$ becomes large. 
As discussed in Appendix \ref{app:2DIsing}, the cumulants of the energy distribution can be found from the free energy $F_N$ 
through the simple relation
\bea \label{cumuli}
\kappa_k = (-1)^{k-1} \frac{\partial^k(\beta F_N)}{\partial \beta^k}.
\eea
Moreover, for the 2D Ising model, an exact expression for  $F_N$, for finite $N$, is available~\cite{Baxter}. 
Using these facts, we show in Appendix \ref{app:2DIsing} that $\kappa_2 \propto N \ln N$~\cite{Fisher_1967, Izmailian_2002}, while $\kappa_k \propto N^{k/2}$, for $k\geq 3$ (these results are also illustrated in Fig.~\ref{fig:IsingGrid}(f)-(h)). 
As a consequence, we therefore have that
\bea
\frac{\kappa_k^{1/k}}{\kappa_2^{1/2}} \propto \ln^{-1/2} N,
\eea
showing that the higher-order cumulants do indeed become negligible as compared to $\kappa_2 = \var(H)$; i.e., the distribution tends to a Gaussian as $N \to \infty$. 

Note that the previous discussion refers to the scaling at the vicinity of the critical point. Away from it, due to the extensivity and analyticity of $F_N$ in the limit of $N \to \infty$, we simply have from Eq.~\eqref{cumuli} that $\kappa_k \propto N$ for $k \geq 1$. Hence, $\kappa_k^{1/k} / \kappa_2^{1/2} \propto N^{- \frac{k - 2}{2 k}}$, i.e., away from criticality, $q(E)$ approaches its Gaussian limit polynomially, as compared to the slow logarithmic convergence at criticality.


\section{Probe-based measurements}
\label{sec:noneqth}

Whereas the previous section allowed for arbitrary global measurements, in this section, we look at measurement schemes (both idealized and more realistic) which can be realised by the interaction of a probe with the system\kar{, possibly some auxiliary system of arbitrary size,} and the subsequent measurement of the probe alone. We compare the performance of such probe-based measurements to the upper bounds obtained in Section~\ref{sec:optbinning}.

First of all, we observe that the maximal thermometric precision achievable by measuring a $d$-dimensional probe $P$ that has unitarily interacted with the system $S$ in a thermal state $\tau$, \kar{and an auxiliary system $\mathcal{A}$ in some state $\rho_{\mathcal{A}}$,} is the same as the maximal precision of a $d$-outcome measurement on $S$. Here we assume that $d < D$, because otherwise one can simply transfer all of the state of $S$---and the (Fisher) information on $\beta$, $\F$, along with it---to $P$; however, when $d < D$, even the best possible strategy of encoding the state of $S$ into that of $P$ will bear losses. Indeed, if the initial state of $P$ is some $\sigma$, then, whatever the optimal unitary $U$, standard quantum metrology tells us \cite{Giovannetti_2011} that the optimal POVM on
\kar{\begin{equation}
    \sigma' = \Tr_{S \mathcal{A}} \big\{ U (\sigma \otimes \tau \otimes \rho_{\mathcal{A}}) U^\dagger\big\}
\end{equation}}
will have to have $d$ outcomes. On the other hand, the probability distribution generated by any $d$-outcome POVM on $\sigma'$ can also be generated by a $d$-outcome POVM on $S$. Thus, denoting the quantum Fisher information of $\sigma'$ on $\beta$ by $\C^{(P)}$, we have $\C^{(P)} \leq \C$.

To show that $\C^{(P)} = \C$, let us note that $\C$ is delivered by a projective measurement on the system corresponding to some binning $I_1 \cup \, \cdots \, \cup I_d$ which yields measurement statistics $p_\alpha = \sum_{q_j \in I_\alpha} q_j$. Now, let us choose $\sigma = \ket{1} \bra{1}$, so that, in the $\{ \ket{\alpha} \otimes \ket{E_j} \}$ basis, $\sigma \otimes \tau = \mathrm{diag}\big( \vec{\mathbf{q}}, \vec{\mathbf{0}}, ..., \vec{\mathbf{0}}\big)$. Here, $\vec{\mathbf{q}} = (q_1, ..., q_D)$ and $\vec{\mathbf{0}}$ is made of $D$ zeroes. Then, the permutation unitary acting on $\sigma \otimes \tau$ that permutes all the $q_j$'s in $I_2$ from $\vec{\mathbf{q}}$ with some of the zeroes from the $\vec{\mathbf{0}}$ next to $\vec{\mathbf{q}}$, all the $q_j$'s in $I_3$ with zeroes from the second $\vec{\mathbf{0}}$, \textit{etc.}, will render $\sigma' = \mathrm{diag} (p_1, ..., p_d)$. And this distribution will produce a $\C^{(P)}$ that is ~$= \C$. \kar{Note that, to show that $\C^{(P)}$ can be made equal to $\C$, there was no need to involve any auxiliary systems.}

For the case of $d = 2$, where the optimal POVM on $S$ is defined by the bins $I_1 = \{E_j: \; E_j < b \}$ and $I_2 = \{E_j: \; E_j \geq b \}$, with $b$ being the boundary, we will now show that such a permutation can be generated by the quantum-optics-inspired Hamiltonian
\begin{equation}
\begin{aligned}
    H = 
    & \sum_{k = 1}^D E_k \ket{E_k} \bra{E_k} + b \ket{\!\Uparrow} \bra{\Uparrow \!}
    \\
    & + \lambda \! \sum_{E_k \leq E_D - b} \! \left(\ket{\!\Downarrow} \bra{\Uparrow \!} \otimes \ket{E_k + b} \bra{E_k} + \mathrm{H.c.} \right) ,
    \label{eq:optint}
\end{aligned}
\end{equation}
where $\ket{\!\Downarrow}$ and $\ket{\!\Uparrow}$ are the eigenstates of the probe spin, with the corresponding eigenvalues $E_{\Downarrow} = 0$ and $E_{\Uparrow} = b$.
\gtl{This Hamiltonian may not be easily realizable in practice. However, the point is that, as we will show, it is guaranteed to provide the best possible precision using a two-level probe. This can then be used as a benchmark to compare against when using other interactions.}
Furthermore, we take the system's ground state to be at energy $E_1 = 0$, and since ultimately we are going to be interested in the case where the system's energy spectrum is effectively continuous, we will also assume that $\ket{E_k + b}$ is a valid eigenstate.

Let us initialize the system and the probe in the joint state $\rho(0) = \ket{\!\Downarrow} \bra{\Downarrow \!} \otimes \tau$. The simplest way to characterize its evolution under $H$ is to describe how the pure states comprising it, $\ket{\Psi_j (0)} = \ket{\!\Downarrow} \otimes \ket{E_j}$, evolve under $H$. It is easy to show that, in the interaction picture (labelled by the superscript $\mathrm{I}$), $\ket{\Psi^{\mathrm{I}}_j (t)} = \ket{\!\Downarrow} \otimes \ket{E_j}$ for $E_j < b$, and
\bea \nonumber
\ket{\Psi^{\mathrm{I}}_j (t)} = \cos (\lambda t) \ket{\!\Downarrow} \otimes \ket{E_j} -i \sin (\lambda t) \ket{\!\Uparrow} \otimes \ket{E_j - b}
\eea
for $E_j \geq b$. Thus, transitioning back to the Schr\"{o}dinger picture, from $\rho(t) = \sum_j \frac{1}{Z} e^{-\beta E_j} \ket{\Psi_j(t)} \bra{\Psi_j(t)}$ we find the probability of finding the probe qubit in the spin-up state, $\mathbb{P}_{\Uparrow}$, when measuring it at the moment of time $t$, to be
\begin{equation}
\mathbb{P}_{\Uparrow} (t) = \sin^2 (\lambda t) \sum_{E_j \geq b} \frac{e^{-\beta E_j}}{Z} .
\end{equation}
Hence, for $t_{\mathrm{meas}} = \pi/(2 \lambda)$, the ideal projective measurement of the probe qubit's spin produces a probability distribution identical to that produced by the ideal binary measurement of the system corresponding to the binning $I_1 \cup I_2$.

Note that realizing this idealized scheme experimentally is far from being straightforward; we study a specific model realization in the next subsection, describing both its capabilities and limitations.

\subsection{Jaynes–Cummings model}

As a specific illustration of quantum probe-based thermometry, we consider an experimentally relevant system consisting of a superfluid Bose-Einstein condensate (BEC) reservoir in a shallow confining trap interacting with an atomic quantum dot \cite{Recati_2005}.
Generally, the physics of this system is well-captured by a spin-boson model, in which the atomic quantum dot interacts with the phononic excitations of the BEC superfluid.
Given suitably engineered boundary conditions, the spectral density will be such that a quantum dot with frequency $\omega_{d}$ will in effect couple only to the phonon modes which comes closest to be resonant with the quantum dot frequency (for simplicity we suppose all relevant phonon modes have the same frequency $\omega_{a}$).

\begin{figure}
    \centering
    \includegraphics[width=0.45\textwidth]{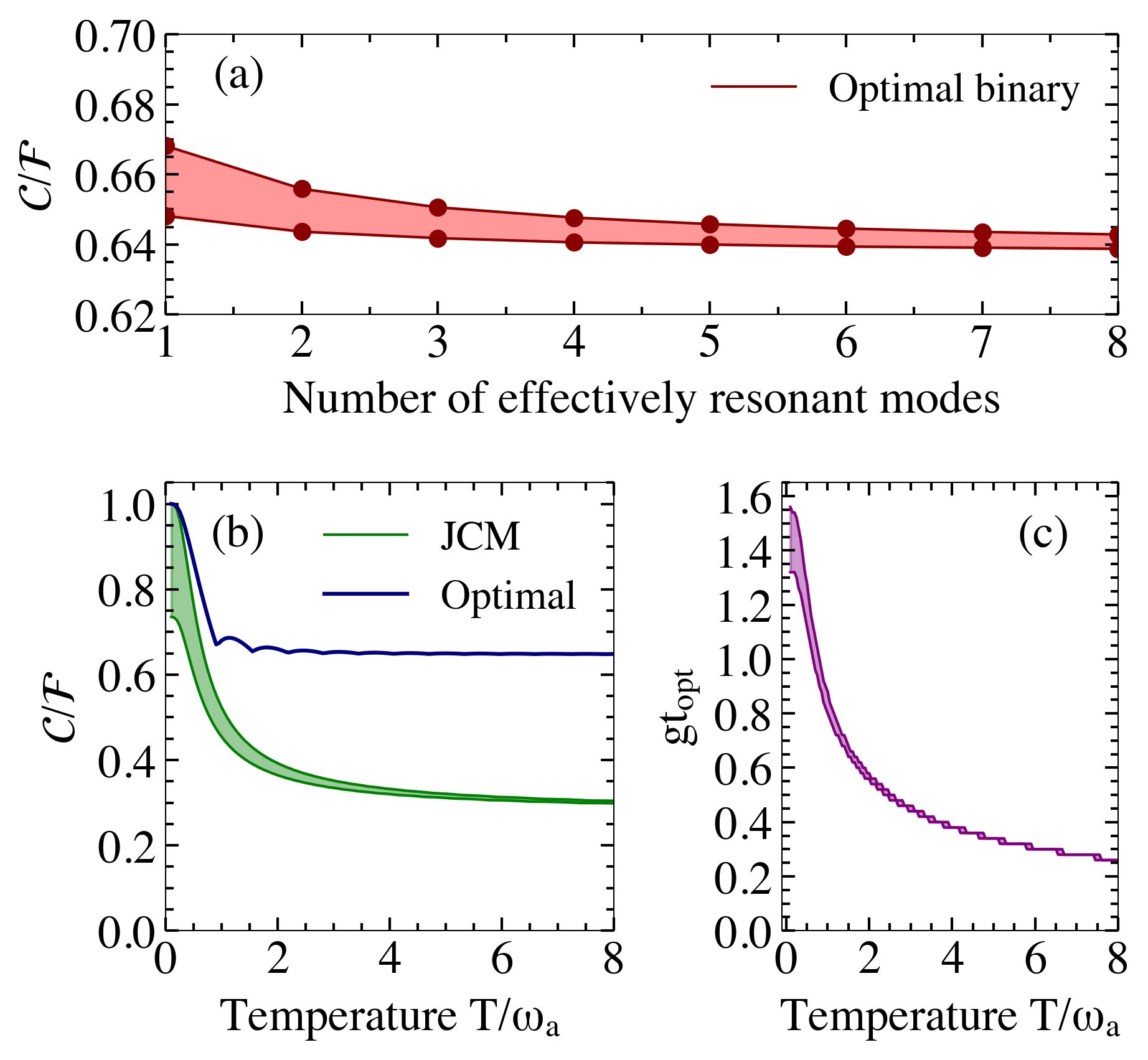}
    \caption{(a) Optimal binary-outcome thermometry on a collection of bosonic modes of frequency $\omega_{a}$ as a function of the number of oscillators.
    The top line corresponds to $\beta\omega_{a} = 0.7$, the bottom line to $\beta\omega_{a} = 0.1$ and intermediate temperatures are contained within the shaded area.
    (b) Comparison between the optimal binary measurement strategy, and the two-level Jaynes-Cummings model probe, optimized over the measurement time $gt$, in the single oscillator case.
    The shaded area represents the range $\delta/g = [0,1.2]$ of the detuning.
    (c) Optimal measurement time ($gt_{opt}$) as a function of temperature for the Jaynes-Cummings model.}
    \label{fig:jcm}
\end{figure}

We can then ask the question of how well one can estimate the BEC temperature by measurements on the quantum dot probe.
In Fig.~\ref{fig:jcm}(a) we plot the coarse-grained vs thermal Fisher information ratio for the optimal binary measurement within the effectively resonant subspace.
The ratio is given as a function of the number of effectively resonant modes, which would be expected to increase in proportion to the width of the spectral density.
From the figure we see that the ratio approaches a value of 0.64 as the number of modes increase.
This provides an optimal value against which to compare specific binary measurement strategies.
Interestingly we see that the obtained ratio agrees with the optimal ratio found for a binary measurement on a system described by a Gaussian density of states.

If we now consider the specific case where the effectively resonant subspace consist of a single phononic mode,
and furthermore make a rotating wave approximation, the resulting system is modelled by the paradigmatic Jaynes-Cummings Hamiltonian~\cite{Langford_2017}:
\begin{equation}
    H = \omega_{d} \sigma^{\dagger}\sigma + \omega_{a} a^{\dagger} a + g \left( \sigma^{\dagger}a + \sigma a^{\dagger} \right) ,
    \label{JCh}
\end{equation}
where $a,a^{\dagger}$ are the creation and annihilation operators of the bosonic cavity mode, $g$ is the coupling strength,
and $\sigma = \ket{g}\!\bra{e}$ where the excited and ground states of the quantum dot are denoted by $\ket{e}$ and $\ket{g}$ respectively.
Experimental work has shown how such models arise for specific thermometry protocols~\cite{Sabin_2014},
and our results makes it possible to evaluate how close such a strategy is to being optimal.

Consider a measurement protocol in which the quantum dot is initialized in its ground state.
The quantum dot then evolves jointly with the BEC for a time $t$, after which the probability of finding the quantum dot in its excited state is give by
\begin{equation} \label{eq:jcm_prob}
    \mathbb{P}_{e}(t) = \sum_{n=0}^{\infty} \frac{e^{-\beta \omega_{a} (n+1)}}{Z_{\beta}} \frac{g^{2}(n+1)}{\lambda_{n}^{2}} \sin^{2} \left( \lambda_{n} t \right) ,
\end{equation}
where we have defined $\lambda_{n} = \sqrt{\delta^{2}/4+g^{2}(n+1)}$ in terms of the detuning $\delta = \omega_{d}-\omega_{a}$.
From this probability we can compute the coarse-grained Fisher information, see also \cite{Brunelli_2011}.
In Fig.~\ref{fig:jcm}(b) we show the ratio of the coarse-grained Fisher information computed from Eq.~\eqref{eq:jcm_prob} and optimized over the measurement time at each temperature,
with the thermal Fisher information of the phononic mode itself.
Inspection of the results show that the probe-based measurement gives a Fisher information which never falls below 45\% of the optimal binary measurement strategy.
In Fig.~\ref{fig:jcm}(c) we plot the optimal measurement time as a function of temperature,
and we observe an inverse relationship between the optimal measurement time and the temperature.
Notice that as the temperature approach absolute zero, the optimal measurement time for zero detuning approach $\pi/2g$ in agreement with known results~\cite{Brunelli_2011}.

It is interesting to note that similar considerations for the Fisher information have been obtained for temperature measurements of micromechanical resonators via a qubit probe, whose interaction can also be described by the Jaynes-Cummings \eqref{JCh}  \cite{Brunelli_2011}; and  more general interaction Hamiltonians, either dropping the rotating wave approximation or considering interactions far off resonance, have also been considered  \cite{Brunelli_2012, Higgins_2013}. In all such cases, our considerations provide upper bounds on the maximal precision estimation with a qubit probe, as shown in Fig. \ref{fig:jcm}. Indeed, the strength of our bounds is that they apply to arbitrary nonequilibrium strategies.




\section{Conclusions}

We considered the precision limits on temperature estimation when having access to coarse-grained measurements which have at most $d$ outcomes. 
Using tools from signal processing, we derived the structure of the optimal POVM measurement. These abstract considerations have been applied to two physically relevant scenarios: temperature measurements of many-body systems and nonequilibrium thermometry.

For many-body systems, we considered spin lattices, both away from and at criticality, and found that the Fisher information $\mathcal{C}$ can grow extensively with the system size even when $d$ does not. In particular, for $d=2$, we obtain that it is in principle possible that $\mathcal{C}/\mathcal{F} \approx 2/\pi$ in the thermodynamic limit $N \rightarrow \infty $ even for systems at criticality.  
While this will decrease for realistic strategies where the POVM are smoothed out (see the discussion in Sec.~\ref{sec:Gremarks} and Fig.~\ref{fig:MessyBins} specifically), we expect that the extensive scaling will be preserved as long as the binary measurement can distinguish system energies that are $O(\sqrt{\var(H)})$ apart (see the discussion on displaced boundary in Sections \ref{sec:gaussianEdist} and \ref{sec:noncrit}).

Along the way, we also derived new results on the energy distribution of many-body systems in the regime of criticality, which might be of independent interest. These generic considerations were illustrated on the $2$D Ising model, the energy distribution of which becomes well-approximated by a Gaussian distribution except in the tails of the distribution. We expect more pronounced non-Gaussian features in the energy distribution of other critical models, which we leave as interesting future research.


For nonequilibrium thermometry,
we used our results to devise the optimal probe-system interaction and interrogation time, thus providing general guidelines on the design of optimal nonequilibrium thermometry strategies. This result also provides an upper bound on specific experimentally motivated protocols. This was illustrated for a temperature measurement of a Bose-Einstein condensate through a quantum dot via a Jaynes-Cummings interaction \cite{Brunelli_2011, Brunelli_2012}. It remains an exciting open question to find a realistic implementation of the optimal probe-sample interaction \eqref{eq:optint}.

\jon{Lastly, in this work, we focused on asymptotic metrology, where one has access to full measurement statistics and can possibly run the experiment many times. This may not always be feasible in practice, and the Fisher information may then no longer be an adequate precision quantifier. In such cases, alternative approaches are needed, such as global Bayesian estimation \cite{Rubio_2020}. Analyzing the effect of coarse-graining in such situations is another interesting research direction.}


\acknowledgments
\mar{We sincerely thank Raffaele Salvia for 
pointing us to Ref.~\cite{Chakravarty_1985}, from which it follows that  consecutive binnings are optimal, as previously conjectured. We also gratefully acknowledge fruitful discussions with Raam Uzdin.}
MRJ and JBB are supported by the Independent Research Fund Denmark.
GTL acknowledges the financial support of the São Paulo Funding Agency FAPESP (Grants No. 2017/50304-7, 2017/07973-5 and No. 2018/12813-0) and the Brazilian funding agency CNPq (Grant No. INCT-IQ 246569/2014-0). M. P.-L. acknowledges funding from Swiss National Science Foundation (Ambizione  PZ00P2-186067).
The authors acknowledge the International Centre for Theoretical Physics, in Trieste, where part of this work was developed, for both the financial support and hospitality.

\bibliographystyle{prxK}
\bibliography{references}

\newpage
\appendix
\onecolumngrid

\section{Optimality of consecutive binning}
\label{app:binning}

\kar{In this appendix, we apply Theorem 1 of Ref.~\cite{Chakravarty_1985} to prove that the sets $I_\alpha$ maximizing the coarse-grained Fisher information $\C$ can be chosen to be consecutive. First, we will state the mentioned theorem, and then we will show how it applies to our problem.}
\kar{
\begin{theorem} \label{th:chakra}
    (Theorem 1 in Ref.~\cite{Chakravarty_1985}) Let $a_1$, ..., $a_D$ and $b_1$, ..., $b_D$ be real numbers such that either $a_i \geq 0$ $\forall i$ or $b_i \geq 0$ $\forall i$ and
    \bea \label{condTh1}
    \frac{a_1}{b_1} \leq \cdots \leq \frac{a_D}{b_D}.
    \eea
    If some $b_i = 0$, then $a_i / b_i$ is defined as $\mathrm{sgn}(a_i) \times \infty$ when $a_i \neq 0$. When $a_i$ is also $0$, $a_i / b_i$ is defined arbitrarily to satisfy Eq.~\eqref{condTh1}. Next, for a partitioning of the set $\{i\}_{i = 1}^D$ into $d$ nonoverlapping subsets $\{I_\alpha\}_{\alpha = 1}^d$, $\bigcup_{\alpha = 1}^d I_\alpha = \{i\}_{i = 1}^D$, define $a_{I_\alpha} := \sum_{i \in I_\alpha} a_i$ and $b_{I_\alpha} := \sum_{i \in I_\alpha} b_i$. Then, the maximum of an objective function
    \bea
    h(a_{I_1}, b_{I_1}, ..., a_{I_d}, b_{I_d})
    \eea
    over all possible partitionings $\{ I_\alpha \}_{\alpha = 1}^d$ is delivered by a partitioning into consecutive subsets if $h$ is a jointly convex function of all of its $2 d$ arguments. (A subset of $\{i\}_{i = 1}^D$ is called consecutive if its elements are consecutive numbers.)
\end{theorem}}

\kar{Returning to our problem, let us assign $a_i = q_i (E_i - \av{H})$ and $b_i = q_i$. Now, we immediately see that $b_i > 0$ for all $i$'s and $a_i / b_i = E_i - \av{H}$ satisfy the condition \eqref{condTh1} in view of Eq.~\eqref{E_ordering}. In terms of $a_{I_\alpha}$ and $b_{I_\alpha}$, our objective function, $\C$ (corresponding to $h$ in the theorem), as defined by Eq.~\eqref{eq:C_def}, reads
\bea
\C(a_{I_1}, b_{I_1}, ..., a_{I_d}, b_{I_d}) = \sum_{\alpha = 1}^d \frac{a_{I_\alpha}^2}{b_{I_\alpha}}.
\eea
In view of the joint convexity of the function $a^2 / b$ \cite{Effros_2009}, we immediately see that $\C(a_{I_1}, b_{I_1}, ..., a_{I_d}, b_{I_d})$ is a jointly convex function of all of its $2 d$ arguments.}

\kar{So, having shown that all the conditions of Theorem~\ref{th:chakra} are met for our problem, we have thus proven that the maximum of $\C$ over all partitions ($\equiv$ nonoverlapping bins) $\{ I_\alpha\}_{\alpha = 1}^d$ is delivered by a consecutive partitioning.}

\kar{Lastly, let us note that the optimality of consecutive partitioning does not preclude the possibility that some other, nonconsecutive partitioning delivers the same maximum. In fact, direct numerical checks show that for our particular case such coincidences do indeed happen.}

\section{NONINTERACTING QUBITS IN THE LARGE $N$ LIMIT}
\label{app:Nqubits}

In the limit of  a large number of qubits, we can approximate the sums in Eqs.~\eqref{N_qubits_p_epsilon} by integrals (using the De Moivre--Laplace theorem), and obtain the approximate probabilities and bin energies
\beaa
    p_\alpha &= \frac{1}{2} \left[ \erf \! \left( \frac{Ns-b_{\alpha - 1}}{\sqrt{2 Nrs}}\right) - \erf \! \left( \frac{Ns - b_\alpha}{\sqrt{2 Nrs}}\right) \right] ,
    \\
    \epsilon_\alpha &= \av{H} - \frac{\sqrt{Nrs}}{\sqrt{2\pi}p_\alpha} \left[ \exp \! \left(-\frac{(b_\alpha - Ns)^{2}}{2Nrs}\right) - \exp \! \left(-\frac{(b_{\alpha - 1} - Ns)^{2}}{2Nrs}\right)\right].
\eeaa

The case of binary measurements, $d=2$,  are  defined by a single bin boundary $b=b_1$. Assuming the system to be at a temperature such that $Ns/r \gg 1$, substituting into Eq.~\eqref{eq:C_def}, we find
\begin{equation} \label{eq:approx}
    \frac{\mathcal{C}}{\F} \approx \frac{2}{\pi} \frac{\exp \! \left(-\frac{(Ns-b)^{2}}{Nrs}\right)}{1 - \erf \! \left( \frac{Ns-b}{\sqrt{2Nrs}} \right)^{2}} .
\end{equation}
The ratio takes its maximum value for $b = Ns$, which is also a solution to Eq.~\eqref{implicit}. In this case we therefore arrive at Eq.~\eqref{NqubCF}. 
This is consistent with Eq.~\eqref{qfidim2} in view of the De Moivre--Laplace theorem that almost straightforwardly states that the energy distribution of $N \gg 1$ thermal identical noninteracting qubits is a discrete Gaussian. Thus, we find that in the large-$N$ limit for a binary measurement, the optimal partition is at the average energy, and a very appealing feature is that such a measurement provides a Fisher information which is already $\frac{2}{\pi} \approx 64\%$ of the maximal possible value, independently of the system size $N$.

\section{PROPORTIONALITY $\mathcal{C} \propto \mathcal{F}$ FOR TWO-BIN MEASUREMENTS ON SHORT TAIL DISTRIBUTIONS} \label{app:propN}

In this appendix we show that a proportionality of the form $\mathcal{C} \propto \F$ is generically true for two-bin measurement, in which the separation of the bins is at the average energy $\av{H}$, for any distribution that decays sufficiently quickly (see Eq.~\eqref{joj1}). Note that, by energy distribution, we understand the $q(E) = \Omega(E) \frac{e^{- \beta E}}{Z}$ [Eq.~\eqref{energdist}], and not just $e^{-\beta E}$. 
Indeed, since $\beta >0$,  $e^{-\beta E}$ must decay monotonically with energy. But $\Omega(E)$ may concentrate energy in different sectors, which can cause the overall distribution $q(E)$ to have arbitrary shapes. 
Indeed, as discussed in Sec.~\ref{sec:manybody}, quite often $q(E)$ will have a Gaussian shape or, more generally, a short tail distribution. 


For a dichotomic measurement, with bin position at $b = \av{H}$, the coarse-grained Fisher information \eqref{eq:C_def} can be written, with some rearrangements, as 
\bea \label{kardol}
\mathcal{C} = \beta^4 \frac{\mathfrak{E}^2}{p_+(1 - p_+)},
\eea
where
\bea
p_+ &=& \int_{\av{H}}^\infty dE~q(E),
\\
\mathfrak{E} &=& \int_{\av{H}}^\infty dE~(E - \av{H})~q(E) = \int\limits_0^\infty dE~E~\hat{q}(E).
\eea
Here, to  simplify notation, we also introduced $\hat{q}(E) = q(E+\av{H})$.

Now, we require---and this is the only requirement we impose---that there exists a fixed (i.e., independent on $N$) number $\lambda > 1$ such that
\bea \label{joj1}
\int_{I_3} dE~E^2~\hat{q}(E) \leq A \var{(H)},
\eea
where
\bea \label{joj2}
0 \leq A < \frac{1}{2}
\eea
and
\bea
I_3 = \{ E: \; |E| > \lambda \sqrt{\var{(H)}} \}.
\eea
These conditions are generically expected to be satisfied by unimodal distributions with fast decaying tails; however, unimodality is not a requirement as long as the above requirement is met.

Let us further pick a
\bea \label{joj3}
0 < \lambda_0 < \frac{1}{2}
\eea
and define
\bea \label{eq:i1}
I_1 = \{ E: \; |E| \leq \lambda_0 \sqrt{\var{(H)}} \}.
\eea
We designate the remaining interval as
\bea \label{eq:i2}
I_2 = \{ E: \; \lambda_0 \sqrt{\var{(H)}} < |E| \leq \lambda \sqrt{\var{(H)}} \},
\eea
and we further define
\beaa
p_k &= \int_{I_k} dE~\hat{q}(E), \qquad k = 1,2,3.
\eeaa
Now, by Eq. \eqref{eq:i1},
\bea \label{joj4}
\int_{I_1} dE~E^2~\hat{q}(E) \leq \lambda_0^2 \var{(H)} p_1 = \lambda_0^2 \var{(H)} (1 - p_2 - p_3).
\eea
Then, again, by the definition of Eq. \eqref{eq:i2},
\bea \label{joj5}
\lambda^2 \var{(H)} p_2 \geq \int_{I_2} dE~E^2~ \hat{q}(E) = \var{(H)} - \int_{I_1 \cup I_3} dE~ E^2~\hat{q}(E).
\eea
Invoking Eqs.~\eqref{joj1} and \eqref{joj4}, we rewrite Eq.~\eqref{joj5} as
\bea
\lambda^2 \var{(H)} p_2 \geq (1 - A - \lambda_0^2) \var{(H)} + \lambda_0^2 \var{(H)} p_2,
\eea
which renders
\bea \label{joj6}
p_2 \geq \frac{1 - A - \lambda_0^2}{\lambda^2 - \lambda_0^2}.
\eea
With this, we immediately find that
\bea
\mathfrak{E} = \int_0^\infty dE~E \hat{q}(E) \geq \int_{I_2} dE~E~\hat{q}(E) \geq \lambda_0 \sqrt{\var{(H)}} p_2,
\eea
and thus,
\bea
\mathfrak{E} \geq \lambda_0 \frac{1 - A - \lambda_0^2}{\lambda^2 - \lambda_0^2} \sqrt{\var{(H)}}.
\eea
Since $p_+(1 - p_+) \leq 1/4$, from Eq.~\eqref{kardol}, we have
\bea
\mathcal{C} \geq \Xi \beta^4 \var{(H)},
\eea
where
\bea
\Xi = 4 \lambda_0^2 \left(\frac{1 - A - \lambda_0^2}{\lambda^2 - \lambda_0^2}\right)^2, \qquad 0 < \Xi < 1.
\eea
Keeping in mind Eq.~\eqref{thermal_fisher}, we thus have
\bea
\Xi \F \leq \C \leq \F.
\eea

Note that the bound above, the purpose of which is solely to prove the scaling of $\mathcal{C}$, may be rather loose, and it is of course desirable to find tighter bounds. We leave this task for future work.

\section{EXAMPLE OF A SMALL FISHER INFORMATION} \label{app:smallC}

In this appendix we construct an example for which the scaling $\mathcal{C} \propto \mathcal{F}$, found in several examples throughout the main text, breaks down. Albeit rather artificial, the goal of this construction is to illustrate that, although occurring generically, the relation $\mathcal{C} \propto \mathcal{F}$ is not universal.

To construct our example, let us take $N$ two-level systems, 
the total Hamiltonian of which is such that the density of states has four sharp peaks, approaching
\bea
\Omega_N(E) = \vartheta_{N, \epsilon} \delta(E - \epsilon) + \vartheta_{N, -\epsilon} \delta(E + \epsilon) + \vartheta_{N, t N} \delta(E - t N) + \vartheta_{N, - t N} \delta(E + t N)
\eea
as $N \to \infty$, where $t > 0$ and $\epsilon > 0$ are some fixed values of energy and the quantities $\vartheta_{N, \pm \epsilon}$ and $\vartheta_{N, \pm t N}$ are also positive. By definition, $\int d E \Omega_N(E) = 2^N$, so
\bea \label{norma}
\vartheta_{N, \epsilon} + \vartheta_{N, - \epsilon} + \vartheta_{N, t N} + \vartheta_{N, - t N} = 2^N.
\eea
We emphasize that the distributions are discrete and become continuous in the $N \to \infty$ limit, and the delta functions above are also to be understood as sharp peaks that approach the delta function in the $N \to \infty$ limit.

Finally, we require that the effectively discrete energy distribution $q_N(E) = \frac{1}{Z} e^{- \beta E} \Omega_N(E)$ [cf. Eq.~\eqref{energdist}] is
\begin{align}
q_{N, E} = \left\{
\begin{array}{ll}
\frac{1}{2} - \frac{1}{N}, & \quad E = \pm \epsilon \\
\\
\frac{1}{N}, & \quad E = \pm t N
\end{array}
\right.,
\end{align}
which corresponds to the following choice of the quantities $\vartheta_{N, E}$:
\bea
\vartheta_{N, \pm \epsilon} &=& \Big( \frac{1}{2} - \frac{1}{N} \Big) e^{\pm \beta \epsilon} Z,
\\
\vartheta_{N, \pm t N} &=& \frac{1}{N} e^{\pm \beta t N} Z,
\eea
where, according to Eq.~\eqref{norma}, $Z$ should be
\bea
Z = \frac{2^N}{\frac{2}{N} \cosh(\beta t N) + \left( 1 - \frac{2}{N} \right) \cosh(\beta \epsilon)}.
\eea

Now, it is straightforward to see that the average energy corresponding to $q_{N, E}$ is zero. Thus, the thermal Fisher information will be
\begin{equation}
\F = \beta^4 \sum_{E = \pm \epsilon, \pm t N} q_{N, E} E^2 = \beta^4 \left[2 t^2 N + \left(1 - \frac{2}{N}\right) \epsilon^2 \right].
\end{equation}
Furthermore, considering two-outcome measurements with the optimal boundary being at the average energy (in this case $0$), the bin probabilities will be $p_1 = p_2 = 1/2$ [cf. Eq.~\eqref{pk_def}] and the bin energies [cf. Eq.~\eqref{epsilon_k}] will be
\begin{equation}
\epsilon_1 = -\epsilon_2 = 2  \bigg[ t + \epsilon \bigg(\frac{1}{2} - \frac{1}{N}\bigg)\bigg],
\end{equation}
hence, according to Eq.~\eqref{eq:C_def},
\begin{equation}
\mathcal{C} = 4 \beta^4 \bigg[t + \epsilon \bigg(\frac{1}{2} - \frac{1}{N}\bigg) \bigg]^2.
\end{equation}
Thus,
\begin{equation}
\frac{\mathcal{C}}{\F} = \frac{4 \left[t + \epsilon \left(\frac{1}{2} - \frac{1}{N}\right) \right]^2}{2 t^2 N + \left(1 - \frac{2}{N}\right) \epsilon^2},
\end{equation}
which, for fixed $\epsilon$ and $t$, scales $\propto 1/N$, which strongly breaks the $\C \propto \F$ relation.

\section{PROOF OF EQ.~\eqref{qfidim2}} \label{app:many}

To prove Eq.~\eqref{qfidim2}, we first state the quantum Berry-Esseen theorem proven in \cite{Brandao_2015, Brandao_2015G} for lattices with finite-range interactions. The theorem requires the following two assumptions about the state of the system $\rho$:
\begin{itemize}
	\item i) The state has exponentially-decaying correlations: for arbitrary regions $\mathcal{X},\mathcal{Y}$ separated by a distance $l$ in the lattice, and some constant $\xi$
	\begin{equation} \label{expodec}
	\max_{X\in \mathcal{X},Y\in \mathcal{Y}} \left | \frac{\langle X \otimes Y \rangle - \langle X \rangle \langle Y \rangle}{\vert \vert X \vert \vert \, \vert \vert Y \vert \vert} \right | \le e^{-l/\xi}.
	\end{equation}
	\item ii) The variance in energy scales with the number of sites as $\var(H) = \av{H^2} - \av{H}^2 = s^2 N$.
\end{itemize}
\kar{The system is an $N$-vertex lattice $\mathcal{L}_N$ (the vertices of correspond to ``particles'') and is described by a locally bounded, finite-range interacting Hamiltonian:
\bea \label{ham1}
H = \sum_{v \in \mathcal{L}_N} H_v.
\eea
Each $H_v$ acts only on vertices the Manhattan distance of which from $v$ is $\leq z$, where $z$ is some fixed natural number that sets the (finite) range of the interactions within the lattice. Lastly, the theorem requires the Hamiltonian to be locally bounded: there exists a constant $\mathfrak{h} > 0$ such that
\bea \label{ham2}
\Vert H_v \Vert \leq \mathfrak{h}, \qquad \forall v \in \mathcal{L}_N,
\eea
where $\Vert \cdot \Vert$ can be chosen to be, e.g., the spectral norm ($\mathfrak{h}$ will of course depend on the norm we choose).}

\begin{theorem} \label{th:berry}
	(Lemma 8 of Ref.~\cite{Brandao_2015})
	Let $\rho$ be a state such that assumptions i) and ii) hold, and with a local Hamiltonian with uniformly bounded local terms, of a system of $N$ particles on a $\mathfrak{d}$-dimensional lattice. Given the cumulative function
	\begin{equation} \label{Jdef}
	J(x)=\sum_{E_i \le x} \bra{E_i}\rho \ket{E_i}
	\end{equation}	
	and the Gaussian cumulative function
	\begin{equation} \label{eq:BEcumul}
	G(x)=\int_{-\infty}^x \frac{\text{d}t}{\sqrt{2\pi \var(H)}} e^{\frac{-(t - \av{H})^2}{2 \var(H)}},
	\end{equation}
	then 
	\begin{equation} \label{BerEss}
	\sup_x |J(x) - G(x)| \le C_0 \frac{\ln^{2 \mathfrak{d}} N}{s^3 \sqrt{N}},
	\end{equation}
	where $C_0$ is a constant.
\end{theorem}
Crucially, the constant $C_0$ does not depend on system size, but only on parameters such as the range of the Hamiltonian ($z$), the lattice structure, or the correlation length $\xi$.

\kar{Note that we can always set the energy of the ground state of $H$ to be zero. Moreover, keeping in mind Eqs.~\eqref{ham1} and \eqref{ham2}, we have $\Vert H \Vert \leq \sum_{v \in \mathcal{L}_N} \Vert H_v \Vert \leq \mathfrak{h} N$, meaning that the largest $E_i$ is $\leq \mathfrak{h} N$. Thus, the real range of energies in Eq.~\eqref{Jdef} can be summarized as $E_{\min} := \min \{ E_i \} = 0$ and $E_{\max} := \max \{ E_i \} \leq \mathfrak{h} N$, and for any $E \not\in [E_{\min}, E_{\max}]$ one sets $q(E) = 0$, so that integrals with infinite energy ranges are meaningful.}

\kar{We now bound $\mathcal{C}$, assuming  two partitions $I_1 = (-\infty, b]$ and $I_2 = (b, \infty)$ (i.e., $d = 2$). Let us first estimate the probabilities $p_1$ and $p_2 = 1 - p_1$, defined in Eq.~\eqref{pk_def},
using Theorem \ref{th:berry}:
\bea \label{eq:pdos}
p_1 = \sum_{E_i \leq b} q_i = J(b) = G(b) + \varepsilon_N,
\eea
where
\bea \label{raccoon}
\varepsilon_N = O \left(\frac{\ln^{2 \mathfrak{d}} N}{\sqrt{N}} \right).
\eea
Next, we note that, since $\av{H} = \sum_{E_i < b} q_i E_i + \sum_{E_i \geq b} q_i E_i$,
\bea \label{ruben}
\sum_{E_i < b} q_i (E_i - \av{H}) = - \sum_{E_i \geq b} q_i (E_i - \av{H}).
\eea
So, upon introducing
\bea \label{frake}
\mathfrak{E} = \sum_{E_i - b \geq 0} q_i (E_i - \av{H}),
\eea
the bin energies, as given by Eq.~\eqref{epsilon_k}, will read
\bea
\epsilon_1 - \av{H} = - \frac{\mathfrak{E}}{p_1} \qquad \mathrm{and} \qquad \epsilon_2 - \av{H} = \frac{\mathfrak{E}}{p_2}.
\eea
Therefore, according to Eq.~\eqref{eq:C_def}, for the coarse-grained Fisher information, we will obtain
\bea \label{maslenica}
\C = \beta^4 [p_1 (\epsilon_1 - \av{H})^2 + p_2 (\epsilon_2 - \av{H})^2] = \beta^4 \frac{\mathfrak{E}^2}{p_1 (1 - p_1)}.
\eea}

\kar{Now, taking into account Eq.~\eqref{eq:pdos}, we can write
\bea
\frac{1}{p_1 (1 - p_1)} = \frac{1}{G(b) [1 - G(b)]} \big\{1 + O(\varepsilon_N [2 G(b) - 1)] + O(\varepsilon_N^2)\big\}.
\eea
Thus, introducing
\bea \label{blinchik}
\tilde{b} = \frac{b - \av{H}}{\sqrt{2 \var(H)}},
\eea
and noticing that
\bea
G(b) = \frac{1}{2} + \frac{1}{2} \erf(\tilde{b}),
\eea
similarly to Eq.~\eqref{maslenica}, we can write
\bea \label{zherekh}
\C = \beta^4 \frac{4 \mathfrak{E}^2}{1 - \erf^2 (\tilde{b})} \big[1 + O(\varepsilon_N \erf(\tilde{b})) + O(\varepsilon_N^2)\big].
\eea}

\kar{Turning to $\mathfrak{E}$, let us rewrite Eq.~\eqref{frake} as
\begin{align} \label{hystrix}
\mathfrak{E} = \sum_{E_i - b \geq 0} q_i (E_i - \av{H}) = \sum_{l=0}^{\infty} \; \sum_{ \Lambda l \leq E_i - b < \Lambda (l+1)} q_i (E_i - \av{H}),
\end{align}
where $\Lambda > 0$ is an arbitrary constant. With this, by choosing an arbitrary natural number $R$, we can lower-bound $\mathfrak{E}$ as
\bea \nonumber
\mathfrak{E} = \sum_{l=0}^{\infty} \; \sum_{ \Lambda l \leq E_i - b < \Lambda (l+1)} q_i (E_i - \av{H}) \geq \sum_{l=0}^{R} \; \sum_{ \Lambda l \leq E_i - b < \Lambda (l+1)} q_i (E_i - \av{H}) \geq \sum_{l=0}^{R} (\Lambda l + b - \av{H}) \sum_{ \Lambda l \leq E_i - b < \Lambda (l+1)} q_i, ~~
\eea
which, upon introducing
\bea \label{kinkajou}
\widetilde{\Lambda} = \frac{\Lambda}{\sqrt{2 \var(H)}},
\eea
will read
\bea \label{marmot}
\mathfrak{E} \geq \sqrt{2 \var(H)} \; \sum_{l = 0}^R \, (\widetilde{\Lambda} l + \tilde{b}) \sum_{\Lambda l \leq E_i - b \leq \Lambda (l + 1)} q_i.
\eea}

\kar{Next, using Theorem \ref{th:berry}, we arrive at the following estimate:
\beaa \label{eq:app3}
\sum_{ \Lambda l \leq E_i - b < \Lambda (l+1)} q_i &= J(b + \Lambda (l+1)) - J(b + \Lambda l) = G(b + \Lambda (l+1)) - G(b + \Lambda l) + \varepsilon_N
\\
&= \int_{b - \av{H} + \Lambda l}^{b - \av{H} + \Lambda (l+1)} \frac{d t}{\sqrt{2 \pi \var(H)}} e^{-\frac{t^2}{2 \var(H)}} + \varepsilon_N = \frac{1}{\sqrt{\pi}} \int_{\tilde{b} + \widetilde{\Lambda} l}^{\tilde{b} + \widetilde{\Lambda} (l + 1)} d t \, e^{-t^2} + \varepsilon_N.
\eeaa
Substituting Eq.~\eqref{eq:app3} into Eq.~\eqref{marmot}, we obtain
\beaa \label{wombat}
\mathfrak{E} &\geq \sqrt{\frac{2 \var(H)}{\pi}} \sum_{l = 0}^R (\tilde{b} + \widetilde{\Lambda} l) \int_{\tilde{b} + \widetilde{\Lambda} l}^{\tilde{b} + \widetilde{\Lambda} (l + 1)} d t \, e^{-t^2} + \sqrt{\var(H)} \, O \big[ (R \tilde{b} + R^2 \widetilde{\Lambda}) \varepsilon_N \big]
\\
&\geq \sqrt{\frac{2 \var(H)}{\pi}} \int_{\tilde{b}}^{\tilde{b} + \widetilde{\Lambda} (R + 1)} d t (t - \widetilde{\Lambda}) e^{-t^2} + \sqrt{\var(H)} \, O \big[ (R \tilde{b} + R^2 \widetilde{\Lambda}) \varepsilon_N \big].
\eeaa
}

\kar{Since $\Lambda$ and $R$ have so far been free, let us choose them such that
\bea \label{skunk}
\widetilde{\Lambda} \to 0 \qquad \mathrm{and} \qquad R \widetilde{\Lambda} \to \infty, \qquad \mathrm{as} \qquad N \to \infty.
\eea
With this choice, we have
\bea \label{pecheni}
\int_{\tilde{b}}^{\tilde{b} + \widetilde{\Lambda} (R + 1)} d t (t - \widetilde{\Lambda}) e^{-t^2} = \int_{\tilde{b}}^\infty d t (t - \widetilde{\Lambda}) e^{-t^2} + O \big(e^{-R^2 \widetilde{\Lambda}^2}\big) = \frac{1}{2} e^{-\tilde{b}^2} + O(\widetilde{\Lambda}) + O \big(e^{-R^2 \widetilde{\Lambda}^2} \big).
\eea
Let us specify our choice of $\widetilde{\Lambda}$ and $R$ to further to
\bea \label{polecat}
\widetilde{\Lambda} = \ln^{-1} N \qquad \mathrm{and} \qquad R = \ln^2 N,
\eea
in full accordance with Eq.~\eqref{skunk} (note that this choice is not unique). With this $\widetilde{\Lambda}$ and $R$, that $e^{-R^2 \widetilde{\Lambda}^2} = e^{-\ln^2 N} = o(\widetilde{\Lambda})$ as $N \to \infty$; therefore, we can absorb the second big O term into the first big O term in Eq.~\eqref{pecheni}. Then, substituting Eq.~\eqref{pecheni} into Eq.~\eqref{wombat}, we arrive at
\bea \label{gorbusha}
\mathfrak{E} &\geq& \frac{1}{2} \sqrt{\frac{2 \var(H)}{\pi}} \big[ e^{-\tilde{b}^2} + O\big(\ln^{-1} N\big) + O\big(\varepsilon_N \tilde{b} \ln^2 N \big) + O\big(\varepsilon_N \ln^3 N\big) \big]
\\ \label{sturg}
&&= \frac{1}{2} \sqrt{\frac{2 \var(H)}{\pi}} \big[ e^{-\tilde{b}^2} + O\big(\ln^{-1} N\big) + O\big(N^{-1/2} \ln^{2 \mathfrak{d} + 3} N\big)  + O\big(\tilde{b} N^{-1/2} \ln^{2 \mathfrak{d} + 2} N \big) \big].
\eea
Since, for sufficiently large $N$, $N^{-1/2} \ln^{2 \mathfrak{d} + 3} N = o\big(\ln^{-1} N\big)$, we can absorb the second big O term into the first big O term in Eq.~\eqref{sturg}, thereby obtaining
\bea \label{sturgeon}
\mathfrak{E} \geq \frac{1}{2} \sqrt{\frac{2 \var(H)}{\pi}} \big[ e^{-\tilde{b}^2} + O\big(\ln^{-1} N\big) + O\big(\tilde{b} N^{-1/2} \ln^{2 \mathfrak{d} + 2} N \big)) \big].
\eea
When $|\tilde{b}| \leq \ln^{4/5} N$, then, by the same logic as above, the second $O$-big term in Eq.~\eqref{sturgeon} can be absorbed into the first $O$-big term, producing
\bea \label{hemp}
\mathfrak{E} \geq \frac{1}{2} \sqrt{\frac{2 \var(H)}{\pi}} \big[ e^{-\tilde{b}^2} + O\big(\ln^{-1} N\big) \big].
\eea}

\kar{In order to estimate $\mathfrak{E}$ in the $|\tilde{b}| \leq \ln^{4/5} N$ range more precisely, let us find an upper-bound for $\mathfrak{E}$ akin to the bound \eqref{hemp}.
To that end, we divide the decomposition in Eq.~\eqref{hystrix} as
\bea \label{otter}
\mathfrak{E} = \sum_{l=0}^{R} \; \sum_{ \Lambda l \leq E_i - b < \Lambda (l+1)} q_i (E_i - \av{H}) + \sum_{l = R+1}^{\infty} \; \sum_{ \Lambda l \leq E_i - b < \Lambda (l+1)} q_i (E_i - \av{H}),
\eea
with $R$ and $\Lambda$ satisfying the conditions in Eq.~\eqref{skunk}.}

\kar{Let us first deal with the second term in Eq.~\eqref{otter}. To do so, recall that our system has exponentially decaying correlations, and therefore Theorem 4.2 of Ref.~\cite{Anshu_2016} applies. It states that, for arbitrary $\mathfrak{d}$-dimensional lattices with exponentially decaying correlations, there exists a constant $\aleph > 0$ such that, whenever $|E - \av{H}| > \aleph \sqrt{\var(H)}$, the function
\bea \label{zabel}
\overline{J}(E) = \left\{\begin{array}{ll}
\int_E^\infty d E' q(E'), & \mathrm{when} \quad E > \av{H} + \aleph \sqrt{\var(H)}
\\ \\
\int_{-\infty}^E d E' q(E'), & \mathrm{when} \quad E < \av{H} - \aleph \sqrt{\var(H)}
\end{array}\right.
\eea
satisfies
\bea \label{habel}
\overline{J}(E) \leq \aleph_0 \exp \bigg(- \widetilde{\aleph} \bigg[ \frac{(E - \av{H})^2}{2 \var(H)} \bigg]^{1/(\mathfrak{d} + 1)} \bigg),
\eea
where $\widetilde{\aleph} > 0$, $\aleph_0 > 0$ are some constants. The bounds provided in Ref.~\cite{Anshu_2016} are a bit tighter, but the bound \eqref{habel} will be sufficient for our needs here.}

\kar{Now, turning to the second sum in Eq.~\eqref{otter}, we write
\beaa \label{badger}
\sum_{l = R+1}^{\infty} \sum_{ \Lambda l \leq E_i - b < \Lambda (l+1)} q_i (E_i - \av{H}) =& \int_{b + (R + 1) \Lambda}^\infty d E (E - \av{H}) q(E) \leq \int_{b + R \Lambda}^\infty d E (E - \av{H}) q(E)
\\
=& (E - \av{H}) \overline{J}(E) \bigg\vert_\infty^{b + R \Lambda} + \int_{b + R \Lambda}^\infty d E \overline{J}(E)
\\
=& (b - \av{H} + R\Lambda) \overline{J}(b + R \Lambda) + \int_{b + R \Lambda}^\infty d E \overline{J}(E),
\eeaa
where in the second line we noted that $q(E) d E = - d \overline{J}(E)$ and performed integration by parts.}

\kar{Now, keeping mind that $-\ln^{4/5} N \leq \tilde{b} \leq \ln^{4/5} N$ and, according to Eq.~\eqref{skunk}, $R \widetilde{\Lambda} = \ln N$, we have that $\tilde{b} + R \widetilde{\Lambda} > 0$ and $\tilde{b} + R \widetilde{\Lambda} \approx \ln N$, which means that $b + R \Lambda - \av{H} \approx \ln N \sqrt{2 \var(H)}$. Hence, inequality \eqref{habel} applies, and therefore we can write
\beaa \label{pelena}
\frac{\sum\limits_{l = R+1}^{\infty} \sum\limits_{ \Lambda l \leq E_i - b < \Lambda (l+1)} q_i (E_i - \av{H})}{\aleph_0 \sqrt{2 \var(H)}} &\leq (\tilde{b} + R \widetilde{\Lambda}) \exp\Big[\! - \widetilde{\aleph} (\tilde{b} + R \widetilde{\Lambda})^{\frac{2}{\mathfrak{d} + 1}}\Big] + \int_{\tilde{b} + R \widetilde{\Lambda}}^\infty d x e^{- \widetilde{\aleph} x^{2/(\mathfrak{d} + 1)}}
\\
&= (\tilde{b} + R \widetilde{\Lambda}) \exp\Big[\! - \widetilde{\aleph} (\tilde{b} + R \widetilde{\Lambda})^{\frac{2}{\mathfrak{d} + 1}}\Big] + \frac{\mathfrak{d} + 1}{2 \widetilde{\aleph}^{\frac{\mathfrak{d} + 1}{2}}} \Gamma \bigg( \frac{\mathfrak{d} +1}{2}, \widetilde{\aleph} (\tilde{b} + R \widetilde{\Lambda})^{\frac{2}{\mathfrak{d} + 1}} \bigg), ~~~~~~
\eeaa
where $\Gamma(x, y)$ is the incomplete gamma function.}

\kar{For $y \gg 1$, $\Gamma(x, y) = y^{x - 1} e^{-y}[1 + O(y^{-1})]$. Therefore, the second term in the last line of Eq.~\eqref{pelena} is
\bea \nonumber
O\Big[(\tilde{b} + R \widetilde{\Lambda})^{\frac{\mathfrak{d} - 1}{\mathfrak{d} + 1}} \exp\big(-\widetilde{\aleph} (\tilde{b} + R \widetilde{\Lambda})^{\frac{2}{\mathfrak{d} + 1}}\big)\Big],
\eea
which means that
\bea \label{blest}
\frac{\sum_{l = R+1}^{\infty} \sum_{ \Lambda l \leq E_i - b < \Lambda (l+1)} q_i (E_i - \av{H})}{\sqrt{2 \var(H)}} = O\Big\{ (\tilde{b} + R \widetilde{\Lambda}) \exp\Big[\! - \widetilde{\aleph} (\tilde{b} + R \widetilde{\Lambda})^{\frac{2}{\mathfrak{d} + 1}}\Big] \Big\}.
\eea
For $N \to \infty$, since $\tilde{b} + R \widetilde{\Lambda} \approx \ln N \to \infty$, the right-hand side of Eq.~\eqref{blest} decays to zero. However, we note that, when $\mathfrak{d} = 3$, it starts becoming small only when $N \gtrsim 10^6$ (which is for $\widetilde{\aleph} = 1$, and for smaller $\widetilde{\aleph}$ it is even slower); for $\mathfrak{d} = 1, 2$, it becomes small much earlier. To obtain a ``safe'' estimate for the big O term in Eq.~\eqref{blest}, let us note that
\bea \label{tsit}
(\tilde{b} + R \widetilde{\Lambda})^2 \exp\Big[\! - \widetilde{\aleph} (\tilde{b} + R \widetilde{\Lambda})^{\frac{2}{\mathfrak{d} + 1}}\Big] = \frac{1}{\widetilde{\aleph}^{\frac{1}{\mathfrak{d} + 1}}} \frac{x^{\mathfrak{d} + 1}}{e^x} \leq \frac{1}{\widetilde{\aleph}^{\frac{1}{\mathfrak{d} + 1}}} \left( \frac{\mathfrak{d} +1}{e} \right)^{\mathfrak{d} + 1} = O(1),
\eea
where we have introduced $x = \widetilde{\aleph} (\tilde{b} + R \widetilde{\Lambda})^{\frac{2}{\mathfrak{d} + 1}}$ for transparency. Equation \eqref{tsit} means that the big O term in Eq.~\eqref{blest} is $O\big[(\tilde{b} + R \widetilde{\Lambda})^{-1}\big] = O \big( \ln^{-1} N \big)$; thus,
\bea \label{nezhdana}
\sum_{l = R+1}^{\infty} \; \sum_{ \Lambda l \leq E_i - b < \Lambda (l+1)} q_i (E_i - \av{H}) \leq \sqrt{\var(H)} O\big( \ln^{-1} N \big).
\eea}

\kar{Now, substituting Eq.~\eqref{nezhdana} into Eq.~\eqref{otter}, we obtain
\beaa \nonumber
\mathfrak{E} =& \sum_{l=0}^{R} \, \sum_{ \Lambda l \leq E_i - b < \Lambda (l+1)} q_i (E_i - \av{H}) + \sqrt{\var(H)} \, O\big(\ln^{-1} N\big)
\\
\leq& \sum_{l=0}^{R} (b- \av{H} + \Lambda (l + 1)) \sum_{ \Lambda l \leq E_i - b < \Lambda (l+1)} q_i + \sqrt{\var(H)} \, O\big(\ln^{-1} N\big),
\eeaa
which, keeping in mind Eqs.~\eqref{blinchik}, \eqref{kinkajou}, and \eqref{eq:app3}, we transform into
\beaa \label{olingo}
\frac{\mathfrak{E}}{\sqrt{2 \var(H)}} \leq& \frac{1}{\sqrt{\pi}} \sum_{l = 0}^R (\tilde{b} + \widetilde{\Lambda} (l + 1)) \int_{\tilde{b} + \widetilde{\Lambda} l}^{\tilde{b} + \widetilde{\Lambda} (l+1)} d t \, e^{-t^2} + O(R^2 \widetilde{\Lambda} \varepsilon_N) + O(R\tilde{b} \varepsilon_N) + O\big(\ln^{-1} N\big)
\\
\leq& \frac{1}{\sqrt{\pi}} \sum_{l = 0}^R \int_{\tilde{b} + \widetilde{\Lambda} l}^{\tilde{b} + \widetilde{\Lambda} (l+1)} d t (t + \widetilde{\Lambda}) e^{-t^2} + O\big(\ln^{-1} N\big) = \frac{1}{\sqrt{\pi}} \int_{\tilde{b}}^{\tilde{b} + \widetilde{\Lambda} (R + 1)} dt (t + \widetilde{\Lambda}) e^{-t^2} + O\big(\ln^{-1} N\big).
\eeaa
Proceeding as in Eq.~\eqref{pecheni} and taking into account Eq.~\eqref{polecat}, we can rewrite Eq.~\eqref{olingo} as
\bea \label{shorzha}
\mathfrak{E} \leq \frac{1}{2} \sqrt{\frac{2 \var(H)}{\pi}} \big[ e^{-\tilde{b}^2} + O\big(\ln^{-1} N\big) \big].
\eea
Thus, when $|\tilde{b}| \leq \ln^{4/5} N$, equations \eqref{hemp} and \eqref{shorzha} show that
\bea \label{salmon}
\mathfrak{E} = \frac{1}{2} \sqrt{\frac{2 \var(H)}{\pi}} \big[ e^{-\tilde{b}^2} + O\big(\ln^{-1} N\big) \big].
\eea}

\kar{Substituted into Eq.~\eqref{zherekh}, Eq.~\eqref{salmon} produces
\bea \label{coelacanth}
\frac{\C}{\F} = \frac{2}{\pi} \frac{\big[ e^{-\tilde{b}^2} + O\big(\ln^{-1} N\big) \big]^2}{1 - \erf^2 (\tilde{b})} \big[1 + O(\varepsilon_N \erf(\tilde{b})) + O(\varepsilon_N^2)\big],
\eea
where we used the fact that $\F = \beta^4 \var(H)$.}

\kar{Note that Eq.~\eqref{coelacanth} resembles Eq.~\eqref{qfidim2}, and shows that the maximum of $\C / \F$ over the boundary position, as quantified by $\tilde{b}$, is reached for some small $\tilde{b}$. To be more specific, Taylor-expanding Eq.~\eqref{coelacanth} around $\tilde{b} = 0$, and considering only $\tilde{b} \ll 1$, we see that
\bea
\frac{\C}{\F} = \frac{2}{\pi} \big[1 - 2 \tilde{b}^2 (1 - 2 / \pi) + O(\ln^{-1} N) + O(\tilde{b}^3) \big].
\eea
This implies that}

\kar{\textbf{(i)}~ Some $\tilde{b}^2 = O\big(\ln^{-1/2} N\big)$ yields the maximum
\bea
\max\limits_{\tilde{b}} \frac{\C}{\F} = \frac{2}{\pi} + O\big(\ln^{-1} N\big),
\eea
and that}

\kar{\textbf{(ii)}~ $\C / \F$ is robust to variations of the bin boundary around its optimal position, in the sense discussed in Sections \ref{sec:gaussianEdist} and \ref{sec:noncrit}.}

\section{ENERGY DISTRIBUTION OF CRITICAL LATTICE SYSTEMS}
\label{app:critdistr}

In this section, we will show that, at a finite-temperature phase transition point, the energy distribution of a translation-invariant, finite-range quantum lattice is Gaussian when the critical exponent $\alpha = 0$ ($\alpha$ is the exponent corresponding to the specific heat; see Sec.~\ref{sec:manybody} and the subsection \ref{app:at} below), albeit with a larger variance as compared to the noncritical case (see Appendix \ref{app:many} and the forthcoming subsection \ref{app:away}). When $\alpha > 0$, we show that the distribution is unimodal with exponentially decaying tails but not in general Gaussian (subsection \ref{app:positivealpha}). The section starts with a derivation of the general formalism for obtaining the energy distributions, then, in Subsection \ref{app:away}, we apply the formalism to the known case of noncritical systems, deriving results consistent with the many-body Berry-Esseen \cite{Brandao_2015, Brandao_2015G} theorem presented in Appendix \ref{app:many}. Finally, in Subsection \ref{app:at} and its two subsubsections, we analyze the energy distribution for critical systems.

In order to understand what the energy distribution near the critical point looks like, we will invoke Lemma 12 from Ref.~\cite{Muller_2015} (see also Theorem III.4.15 of Ref.~\cite{Simon}) stating that the cumulative density of states of a translationally invariant, finite-range lattice in arbitrary spatial dimensions is an exponential of the canonical entropy of the lattice. To cast this in more precise terms, we first have to introduce some notation. Fixing periodic boundary conditions, let $H_N$ denote the translation-invariant Hamiltonian of an $N$-site lattice ($N \gg 1$). Then, for an arbitrary translation-invariant state $\Upsilon_N$ on the lattice, we define the energy and entropy densities as
\bea
u_N(\Upsilon_N) = \frac{1}{N} \Tr (H_N \Upsilon_N) \qquad \mathrm{and} \qquad s_N(\Upsilon_N) = - \frac{1}{N} \Tr (\Upsilon_N \ln \Upsilon_N).
\eea
Now, introducing the free energy density:
\bea \label{freeendens}
f_N(\beta) = - \frac{T}{N} \ln Z_N,
\eea
where
\bea
Z_N = \Tr e^{- \beta H_N},
\eea
we can formulate the variational principle for finite $N$:
\bea \label{var_princ}
\min_{\Upsilon_N} [u_N (\Upsilon_N) - T s_N(\Upsilon_N)] = f_N(\beta) \qquad \mathrm{and} \qquad \arg \min_{\Upsilon_N} [u_N (\Upsilon_N) - T s_N(\Upsilon_N)] = \tau_N(\beta),
\eea
where
\bea
\tau_N(\beta) = \frac{1}{Z_N} e^{- \beta H_N}
\eea
is the Gibbs state. The finite-$N$ principle is read-off straightforwardly from the identity
\bea
\av{H_N}_{\Upsilon_N} - T S(\Upsilon_N) = F_N (\beta) + T S(\Upsilon_N || \tau_N(\beta)),
\eea
where $S(\Upsilon_N) = - \Tr (\Upsilon_N \ln \Upsilon_N)$ is the von Neumann entropy, $S(\Upsilon_N || \tau_N(\beta)) = \Tr ( \Upsilon_N [\ln \Upsilon_N - \ln \tau_N(\beta)])$ is the relative entropy, and $F_N(\beta) = - T \ln Z_N = N f_N (\beta)$ is the free energy of the lattice at temperature $T$.

As can be seen, the finite-$N$ situation always yields a thermal state as the unique solution of the minimization in Eq.~\eqref{var_princ}, and therefore cannot account for second-order phase transitions at finite temperatures. This is of course in accord with the general understanding that finite-temperature phase transitions appear only in the thermodynamic limit ($N \to \infty$). In this limit, the Hilbert space has infinite dimensions, and the simple finite-dimensional argumentation logic breaks down on certain levels. However, as is proven in Refs.~\cite{Araki_1974, Simon}, first of all, the densities
\bea \label{defdefdef}
u(\Upsilon) = \lim_{N \to \infty} u_N(\Upsilon_N), \qquad s(\Upsilon) = \lim_{N \to \infty} s_N(\Upsilon_N), \qquad \textrm{and} \qquad f(\beta) = \lim_{N \to \infty} f_N(\beta)
\eea
exist and the variational principle still holds:
\bea \label{VP_inf}
\inf_{\Upsilon} [u(\Upsilon) - T s(\Upsilon)] = f(\beta),
\eea
where the infimum is sought over the set of translationally invariant states. The infimum is delivered by state(s), which we will call ``equilibrium state(s)'', satisfying the Kubo–Martin–Schwinger (KMS) condition (see, e.g., Refs.~\cite{Araki_1974, Simon} for a definition; we will not go into the details of it since we will not use that definition in what follows) at inverse temperature $\beta$. The states that satisfy the KMS condition for a given $\beta$ generalize the Gibbs state, always coinciding with it when the Hilbert space is finite-dimensional. In infinite dimensions, the KMS state is unique and coincides with the Gibbs state only \textit{at and below} the critical $\beta_c$, if $\beta_c$ is finite. Above $\beta_c$, the KMS state will generally not be unique, with different KMS states representing different phases (see the discussion in Chapter V of Ref.~\cite{Haag}). When $\beta_c = + \infty$, the KMS state is a Gibbs state for arbitrary $\beta < +\infty$; at $\beta_c$, i.e., when the system is in the degenerate ground state, the set of KMS states coincides with the ground eigensubspace. Here, we will only deal with lattices for which $\beta_c$ is finite (that is, only thermal phase transitions).

Defining the minimal and maximal possible energy densities as
\bea
u_{\min} := \lim_{N \to \infty} \frac{\mathrm{min.} \;\; \mathrm{eigenvalue} \;\; \mathrm{of} \;\; H_N}{N} \qquad \mathrm{and} \qquad u_{\max} := \lim_{N \to \infty} \frac{\Tr H_N}{N d^N},
\eea
where $d$ is the local Hilbert-space dimension of a single node of the lattice, we invoke Lemma 9 of Ref.~\cite{Muller_2015} (also proven in Ref.~\cite{Simon}). It states that, for any $u \in (u_{\min}, u_{\max}]$, there exists a unique $\beta = \beta (u)$ for which at least one equilibrium state at temperature $\beta (u)$ yields energy density $u$. Moreover, the entropy density is
\bea \label{entdens}
s(u) = \beta(u) [u - f(\beta(u))],
\eea
and an analogue of the maximum entropy principle holds: $s(u)$ is the highest entropy density among translationally invariant states with energy density $u$.

With this, we are ready to state the result of Lemma 12 of Ref.~\cite{Muller_2015} (which is a clarification and generalization of Theorem III.4.15 of Ref.~\cite{Simon}): if
\bea
Q_N(u) := \# \{\mathrm{eigenvalues} \;\; \mathrm{of} \;\; H_N \;\; \leq u N \},
\eea
then
\bea \label{lemma12}
\lim_{N \to \infty} \frac{\ln Q_N(u)}{N} = s(u).
\eea
Assuming differentiability of $\ln Q_N(u) / N$ with respect to the small parameter $1 / N$, we can write Eq.~\eqref{lemma12} as $\ln Q_N (u) / N = s(u) + O(1 / N)$, or $Q_N (u) = e^{N s(u) + O(1)}$. Furthermore, also assuming differentiability of $s_N (u)$ with respect to $1 / N$, we can write $s_N (u) = s(u) + O(1 / N)$ (this can be rigorously proven for $1$D and $2$D systems even without assuming analyticity \cite{Griffiths_1964}), thereby arriving at
\bea
Q_N (u) = e^{N s_N (u) + O(1)}.
\eea
Now, introducing the density of states (cf. Eq.~\eqref{DOS}),
\bea
\Omega_N(E) = \frac{d Q_N(u)}{N d u},
\eea
where $E = N u$, we obtain from Eq.~\eqref{lemma12}
\bea \label{DDOS}
\Omega_N(E) = e^{N s_N(u) + O(1)}.
\eea
With these, we can write the energy distribution, in the sense of Eq.~\eqref{energdist}, as
\bea \label{dstrb}
q_N(E) = \frac{1}{Z_N} e^{- \beta N [u - T s_N(u)] + O(1)} = e^{\beta N f_N(\beta) - \beta N [u - T s_N(u)] + O(1)},
\eea
where, again, $E = N u$, and, in order to obtain the second equality, we used Eq.~\eqref{freeendens}. The normalization condition thus takes the form
\bea \label{normaliz}
\sum_n \frac{e^{- \beta E_n}}{Z_N} = 1 = \int_{u_{\min}^{(N)}}^{u_{\max}^{(N)}} d(u N) \, e^{\beta N f_N(\beta) - \beta N [u - T s_N(u)] + O(1)}.
\eea

Regardless of what is the exact form of $q_N(E)$, by construction it satisfies
\bea
\int d (u N) \, q_N(u N) \, u &=& u_N(\beta)
\\
\int d (u N) \, q_N(u N) (u - u_N(\beta))^2 &=& \var(u) = \beta^{-2} N^{-1} c_N(\beta). ~~
\eea

Fixing an arbitrary $\beta_0 \leq \beta_c$ we observe that the infimum in Eq.~\eqref{VP_inf} will be given by the Gibbs state at temperature $T_0$ (see the discussion below Eq.~\eqref{VP_inf}), and, since $N$, however large, is finite, the infimum will in fact be a minimum. Therefore, in the vicinity of $u_N(\tau_N(\beta_0))$ (which we will simply call $u_N(\beta_0)$), we can write
\bea \nonumber
u - T_0 s_N(u) = f_N(\beta_0) + \frac{d^2 [u - T_0 s_N(u)]}{d u^2} \bigg|_{u = u_N(\beta_0)} \! \frac{(u - u_N(\beta_0))^2 \!}{2} + \frac{d^3 [u - T_0 s_N(u)]}{d u^3} \bigg|_{u = u_N(\beta_0)} \! \frac{(u - u_N(\beta_0))^3 \!}{6} + \cdots. ~
\\ \label{taylored}
\eea
Now, writing the double derivative in this formula as
\bea
\frac{d^2 [u - T_0 s_N(u)]}{d u^2} = - T_0 \frac{d^2 s_N(u)}{d u^2} = - T_0 \frac{d}{d u} \frac{d s_N(u)}{d u},
\eea
and noting that $\frac{d s_N(u)}{d u} = \beta(u)$, we find that
\bea \label{zozh}
\frac{d^2 [u - T_0 s_N(u)]}{d u^2} = \frac{T_0}{T(u)^2} \frac{d T(u)}{d u} = \frac{T_0}{T(u)^2} \frac{1}{c_N(\beta(u))},
\eea
where $c_N(\beta)$ is the specific heat of the $N$-site lattice at the inverse temperature $\beta$:
\bea
c_N(\beta) := \frac{d u_N(\tau_N(\beta))}{d T}.
\eea
Observing that
\bea \label{blabla}
\beta (u_N(\beta_0)) = \beta_0,
\eea
we thus obtain
\bea \label{K2}
K_2 := \frac{d^2 [u - T_0 s_N(u)]}{d u^2} \bigg|_{u = u_N(\beta_0)} = \frac{\beta_0}{2 c_N(\beta_0)}.
\eea
The cubic term is obtained by differentiating Eq.~\eqref{zozh}:
\bea \label{third}
\frac{d^3 [u - T_0 s_N(u)]}{d u^3} = \frac{T_0 \beta(u)^4}{c_N(\beta(u))^3} \frac{d c_N(\beta)}{d \beta} \bigg|_{\beta = \beta(u)} - \frac{2 T_0 \beta(u)^3}{c_N(\beta(u))^2},
\eea
so that
\bea
K_3 := \frac{d^3 [u - T_0 s_N(u)]}{d u^3} \bigg|_{u = u_N(\beta_0)} = \frac{\beta_0^3}{6 c_N(\beta_0)^3} \frac{d c_N(\beta)}{d \beta} \bigg|_{\beta = \beta_0} - \frac{\beta_0^2}{3 c_N(\beta_0)^2}.
\eea
The higher-order terms in Eq.~\eqref{taylored} ($K_{j \geq 4}$) can be obtained by further differentiating Eq.~\eqref{third}.

Now, separating the $u$-dependent part of the $O(1)$ in Eq.~\eqref{dstrb} as 
\bea
\mathrm{const} + K_1 (u - u_N(\beta_0)) + \sum_{j \geq 2} \frac{\zeta_j}{j!} (u - u_N(\beta_0))^j,
\eea
where $K_1$ and all $\zeta_j$'s are $O(1)$ (since they only depend on intensive quantities), we can write Eq.~\eqref{dstrb} as
\bea \label{qnem}
q_N (E) = K \exp \bigg[K_1 (u - u_N(\beta_0)) - \frac{1}{2} (N \beta_0 K_2 - \zeta_2) (u - u_N(\beta_0))^2 - \sum_{j \geq 3} \frac{1}{j!} (N \beta_0 K_j - \zeta_j) (u - u_N(\beta_0))^j \bigg], ~~
\eea
where, as usual, $E = u N$. $K$ is a $u$-independent quantity that absorbs all $u$-independent quantities; it is nothing but the normalization factor for $q_N(E)$.

\subsection{Away from criticality: $\beta_0 < \beta_c$}
\label{app:away}

When the lattice is away from criticality, $c_N(\beta)$ and its derivatives are finite (i.e., do not scale with $N$). Therefore, $\var(u) \propto \beta^{-2} N^{1}$, meaning that when deviating from the average, $\av{u}$ by $\gg 1$ [but $o(\sqrt{N})$] standard deviations, $u - \av{u}$ remains $o(1)$. Hence, up to rather far into the tails of the distribution, the quadratic term in the exponent in Eq.~\eqref{qnem} dominates the higher-order terms. Neglecting those higher-order terms and combining the linear and quadratic terms, and absorbing the resulting $u$-independent $e^{\frac{K_2^2 c_N(\beta_0)}{2 \beta_0^2 N}}$ into $K$, we find
\bea
q_N(E) = K \exp \bigg[- \frac{\beta_0^2 N}{2 c_N(\beta_0)} (u - \widetilde{u}_N(\beta_0))^2 \bigg],
\eea
where
\bea \label{assym}
\widetilde{u}_N(\beta_0) = u_N(\beta_0) + \frac{K_1 c_N(\beta_0)}{\beta_0^2 N}.
\eea
Noticing that the difference between $\widetilde{u}_N(\beta_0)$ and $u_N(\beta_0)$ is only $\propto 1 / N$, meaning that $\widetilde{\av{H_N}}_{\beta_0} - \av{H_N}_{\beta_0} = O(1) c_N = O(c_N)$ (which is negligible when $N \gg 1$), and rescaling $u$ to $E = N u$, we find that
\bea \label{opilliones}
q_N(E) = K \exp \Big[ - \frac{(E - \av{H_N}_{\beta_0} + O(c_N))^2}{2 \var(H_N)_{\beta_0}} \Big].
\eea

Our result in this subsection thus complements the many-body Berry-Esseen theorem (Theorem \ref{th:berry} in Appendix \ref{app:many}).
In particular, the fact that the tails of $q_N(E)$ decay exponentially [Eq.~\eqref{opilliones}] cannot be directly deduced from the many-body Berry-Esseen theorem; in this sense, our result directly connects to Theorem 4.2 of Ref.~\cite{Anshu_2016} (discussed in Appendix \ref{app:many}) by strengthening it for the particular case of translationally invariant lattices. 

Equation \eqref{opilliones} reflects the ``common wisdom'' that the thermal state is located in a typical subset of energy levels of width $O(\sqrt{\var(H_N)_{\beta_0}}) = O(\sqrt{N})$, centered at the average energy; the energy levels within the typical subset have approximately equal probabilities, so that the entropy of the state is essentially the $\log$ of the number of the energy levels in the subset (which we see by invoking Eq.~\eqref{DDOS} and noting that the number of energy levels in the typical subset is $\propto \sqrt{N} e^{N s_N(\beta_0)}$).

\subsection{At criticality: $\beta_0 = \beta_c$}
\label{app:at}

The specific heat of a critical lattice diverges with $N$, and even more divergent are its derivatives. This necessitates a careful bookkeeping of all the terms in the series in Eq.~\eqref{qnem}. In order to do so, we will need to find how the specific heat and its temperature derivatives scale at criticality.

First, we recall that, as the system approaches the critical temperature, with the approach being parameterized by
\bea \label{ttdef}
t = \frac{\beta - \beta_c}{\beta_c} \ll 1,
\eea
the specific heat and correlation length ($\xi$) in the thermodynamic limit scale as 
\bea \label{critex1}
c_\infty(\beta) &\propto& |t|^{-\alpha},
\\
\xi(\beta) &\propto& |t|^{-\nu},
\eea
where $1 > \alpha \geq 0$ and $\nu > 0$ are the corresponding critical exponents. When $\alpha = 0$ (e.g., in 2D Ising model),
\bea \label{critex2}
c_\infty(\beta) \propto \ln |t|^{-1}.
\eea

When $N$ is finite, neither $\xi(\beta)$ nor $c_N(\beta)$ can diverge as $t \to 0$. In this case, we note that, since the correlation length diverges in the thermodynamic limit, when the lattice is large but finite, it will simply become proportional to the size of the lattice (for spatial dimension $< 4$) \cite{Brezin_1982}. Therefore, in $\mathfrak{d}$ spatial dimensions,
\bea \label{critex3}
\xi(\beta_c) \propto N^{1/\mathfrak{d}}.
\eea
On the other hand, Eq.~\eqref{critex1} suggests that $c_N \propto \xi^{\alpha / \nu}$. Therefore,
\bea \label{critex4}
c_N(\beta_c) \propto N^{\frac{\alpha}{\nu \mathfrak{d}}},
\eea
and when $\alpha = 0$,
\bea \label{critex5}
c_N(\beta_c) \propto \ln N;
\eea
for the proof of this in the case of 2D Ising model, see Ref.~\cite{Izmailian_2002}.

By the same logic, for any $\alpha \geq 0$, we have
\bea \label{critex6}
\frac{d^j c_N(\beta)}{d \beta^j} \bigg|_{\beta = \beta_c} \propto \beta_c^{-j} N^{\frac{\alpha + j}{\nu \mathfrak{d}}},
\eea
since, in view of Eqs.~\eqref{critex1} and \eqref{critex2}, $\frac{d^j c_\infty(\beta)}{d \beta^j} \propto \beta^{-j} |t|^{-\alpha - j}$.

Looking into the structure of Eq.~\eqref{third} and its derivatives with respect to $u$, we see that the term 
\bea
\frac{\beta_c^{2 j - 3}}{c_N(\beta_c)^j} \frac{d^{j - 2} c_N(\beta)}{d \beta^{j - 2}} \bigg|_{\beta = \beta_c}
\eea
is dominant in $K_j$, $j \geq 3$. Therefore,
\bea \label{higherorder1}
K_j \propto \beta_c^{j - 1} \Bigg\{
\begin{array}{ll}
(\ln N)^{-j} N^{\frac{j - 2}{\nu \mathfrak{d}}}, & \mathrm{when} \quad \alpha = 0
\\
N^{\frac{j(1 - \alpha) + \alpha - 2}{\nu \mathfrak{d}}}, & \mathrm{when} \quad \alpha > 0
\end{array} .
\eea
Taking into account the scaling relation (see, e.g., Ref.~\cite{Fisher_1967})
\bea
\nu \mathfrak{d} = 2 - \alpha,
\eea
we can simplify Eq.~\eqref{higherorder1} to
\bea \label{higherorder2}
K_j \propto \beta_c^{j - 1} \Bigg\{
\begin{array}{ll}
(\ln N)^{-j} N^{\frac{j - 2}{2}}, & \mathrm{when} \quad \alpha = 0
\\
N^{j \frac{1 - \alpha}{2 - \alpha} - 1}, & \mathrm{when} \quad \alpha > 0
\end{array} .
\eea

In order to determine the regime of validity of the Gaussian approximation to $q_N$, we will now compare the $j \geq 3$ terms to the quadratic term in Eq.~\eqref{qnem}:
\bea \label{compar1}
(N \beta_c K_2 -\zeta_2) (u - u_N(\beta_c))^2 \gg (N \beta_c K_j - \zeta_j) (u - u_N(\beta_c))^j.
\eea
Since $N K_2$ and all $N K_j$'s diverge with $N$, in the above inequality the $\zeta$'s, being $O(1)$, are not going to play a role, and therefore, we will omit them.

We will analyze $\alpha = 0$ and $\alpha > 0$ cases separately.

\subsubsection{Energy distribution for $\alpha = 0$}
\label{app:alphazero}

In this case, taking Eq.~\eqref{higherorder2} in to account, Eq.~\eqref{compar1} takes the form
\bea
\beta_c (\ln N)^{-1} (u - u_N(\beta_c))^2 \gg \beta_c^{j - 1} (\ln N)^{-j} N^{\frac{j - 2}{2}} |u - u_N(\beta_c)|^j, \qquad \forall j \geq 3,
\eea
which reduces to
\bea
|u - u_N(\beta_c)| \ll \beta_c^{-1} \frac{(\ln N)^{1 + \frac{1}{j - 2}}}{\sqrt{N}}, \qquad \forall j \geq 3.
\eea
The latter simply means that, as long as
\bea
|u - u_N(\beta_c)| \lesssim T_c \frac{\ln N}{\sqrt{N}},
\eea
the quadratic term in Eq.~\eqref{qnem} will dominate the higher-order terms. Noting that, for $\alpha = 0$, the standard deviation of $u$, $\sqrt{\var(u)}$, is $\propto T_c \frac{\sqrt{\ln N}}{\sqrt{N}}$, we conclude that the energy distribution is Gaussian up until $\propto \sqrt{\ln N}$ standard deviations into the tails.

Just like when away from criticality, the linear term in Eq.~\eqref{qnem} shifts the tip of the distribution function by $\propto T_c c_N$ (see Eq.~\eqref{assym}), implying an asymmetry of the distribution as a whole. However, this shift, being $\propto T_c \ln N$, is $\ll$ than the standard deviation $\var(E) \propto T_c \sqrt{N \ln N}$, meaning that the energy distribution, in the energy range in which it is Gaussian, is close to Eq.~\eqref{eq:BEcumul} (and Eq.~\eqref{eq:gaussian}).

We will quantify the proximity of $q_N(E)$ to a Gaussian more precisely for the classical $2$D Ising model in Appendix~\ref{app:2DIsing}.


\subsubsection{Energy distribution for $1 > \alpha > 0$}
\label{app:positivealpha}

For strictly positive $\alpha$'s, Eq.~\eqref{compar1} becomes
\bea
\beta_c N^{- \frac{\alpha}{2 - \alpha}} (u - u_N(\beta_c))^2 \gg \beta_c^{j - 1} N^{j \frac{1 - \alpha}{2 - \alpha} -1} |u - u_N(\beta_c)|^j,
\eea
which leads to
\bea
|u - u_N(\beta_c)| \ll T_c N^{- \frac{1 - \alpha}{2 - \alpha}}.
\eea
Taking into account that the standard deviation of $u$,
\bea
\sqrt{\var(u)} = T_c \sqrt{\frac{c_N(\beta_c)}{N}} \propto T_c N^{- \frac{1 - \alpha}{2 - \alpha}},
\eea
we see that the situation with Gaussianity here is more tricky than for noncritical lattices or those that are critical but with $\alpha = 0$. We see that, once we depart one standard deviation away from the average, we already find ourselves in a situation where both the quadratic term and the higher-order terms are $O(1)$. However, as long as $|E - \av{H_N}_{\beta_c}| \ll T_c N^{1 / (2 - \alpha)}$, $q_N(E)$ tends to a Gaussian as $N \to \infty$.

Lastly, since $\alpha < 1$, $c_N / N \to 0$ as $N \to \infty$, the shift of the peak of the Gaussian caused by the linear term in Eq.~\eqref{qnem}, being $\propto c_N / N$ (see Eq.~\eqref{assym}), is $\ll$ the variance of $u$, which is $\propto \sqrt{c_N / N}$. Therefore, as in the previous cases, we can neglect that effect, while, of course, keeping in mind that it indicates a certain asymmetry of the overall energy distribution, with the asymmetry becoming the more significant, the higher the $\alpha$.

To sum up, for $|E - \av{H_N}_{\beta_c}| \ll T_c N^{1 / (2 - \alpha)}$, $q_N(E) \propto \exp\left[- \frac{(E - \av{H_N}_{\beta_c})^2}{2 N c_N(\beta_c)}\right]$. That $q_N(E)$ significantly deviates from a Gaussian when $\alpha > 0$ is not surprising in the light of Appendix~\ref{app:2DIsing}.

Furthermore, since $s_N(u)$ is a strictly concave function of $u$ (see Theorem III.4.13 of Ref.~\cite{Simon}), and therefore so is $s_N(u) - \beta_c u$, we see from Eq.~\eqref{dstrb} that $- \frac{1}{N}\ln q_N(E)$ is a strictly monotonically increasing function once one departs away from its unique minimum near $\av{H_N}_{\beta_c}$. The strict convexity of $s_N(u) - \beta_c u$ in particular means that, once $|u - u_N(\beta_c)| = \Theta (1)$, where the big $\mathrm{\Theta}$ is according to standard asymptotic notation, $q_N(N u) = e^{-N \Theta(1)}$. In other words, the distribution $q_N(E)$ is unimodal, with exponentially decaying tails.

\subsection{Classical 2D Ising model at phase transition}
\label{app:2DIsing}

Let us now look into the the most salient example of a finite-temperature phase-transition with $\alpha = 0$---the classical square-lattice $2$D Ising model. To make the inevitably complicated analysis as easy as possible, we choose the model to have symmetric, nearest-neighbour couplings, all equal to $1$ ($J = 1$); be at zero magnetic field; and have periodic boundary conditions on both boundaries. In such a case, the free energy per particle in the thermodynamic limit, $f = \lim_{N \to \infty} \frac{1}{N} F_N = - \frac{\beta^{-1}}{N} \ln Z_N$ (cf. Eq.~\eqref{defdefdef}), is given by \cite{Baxter}
\beaa
\beta f = - \frac{\ln 2}{2}- \ln[\cosh(2 \beta)] + \frac{1}{2 \pi} \int_0^\pi \! d\theta \ln \! \left[ 1 + \sqrt{1 - \iota^2 \cos^2 \theta} \right],
\eeaa
where
\bea
\iota = \frac{2 \sinh(2 \beta)}{\cosh^2 (2 \beta)},
\eea
with the critical temperature being
\bea
\beta_c = \frac{\ln(1 + \sqrt{2})}{2}.
\eea

Differentiating $\beta f (\beta)$ we obtain:
\beaa
\frac{\partial (\beta f)}{\partial \beta} &= \lim_{N \to \infty} \frac{1}{N} \av{H}_\beta,
\\
\frac{\partial^2 (\beta f)}{\partial \beta^2} &= - \lim_{N \to \infty} \frac{1}{N} \av{(H - \av{H}_\beta)^2}_\beta := - \lim_{N \to \infty} \frac{1}{N} E_2 = - \beta^{-2} c(\beta),
\eeaa
where, as before, $c(\beta)$ is the specific heat and
\bea
E_n := \av{(H - \av{H}_\beta)^n}_\beta
\eea
is $n$'th central moment. Now, using the easy-to-derive formula
\bea \label{iterat}
E_{n + 1} = n E_{n - 1} E_2 - \frac{\partial E_n}{\partial \beta}, \quad n \geq 2,
\eea
we immediately obtain
\bea \label{E3thelim}
\frac{\partial^3 (\beta f)}{\partial \beta^3} = \lim_{N \to \infty} \frac{1}{N} \av{(H - \av{H}_\beta)^3}_\beta = \lim_{N \to \infty} \frac{1}{N} E_3.
\eea

\begin{figure}[t!]
\subfloat[]{\includegraphics[width = 0.47 \textwidth]{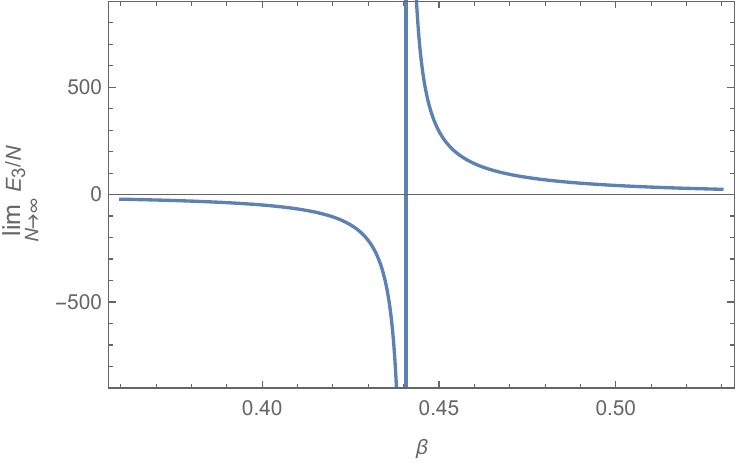}}
\hspace{8 mm}
\subfloat[]{\includegraphics[width = 0.47 \textwidth]{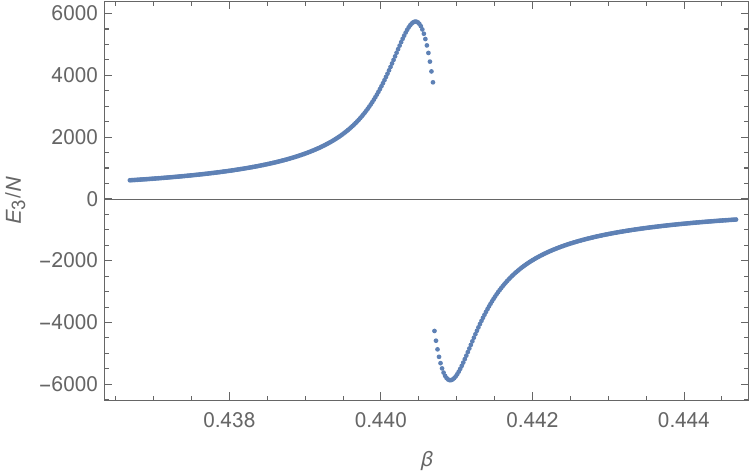}}
\caption{\textbf{(a)} The density of the third moment of energy of the $2$D Ising model in the thermodynamic limit, defined by Eq.~\eqref{E3thelim}, plotted against the inverse temperature $\beta$. \textbf{(b)} The same quantity for $N = 10^6$, calculated using the exact solution of the $2$D Ising model, plotted against $\beta$.}
\label{fig:E3}
\end{figure}

Near the critical temperature \cite{Fisher_1967}
\bea
c(\beta) \propto \ln |b|^{-1},
\eea
which, at criticality, translates into \cite{Izmailian_2002}
\bea
E_2 \propto N \ln N.
\eea
Using $E_3 = -\frac{\partial E_2}{\partial \beta} = 2 N \beta^{-3} c(\beta) - N \beta^{-2} \frac{\partial c(\beta)}{\partial \beta}$, we see that
\bea \label{hoff1}
E_3 \propto N |b|^{-1},
\eea
see Fig.~\ref{fig:E3}(a). This vaguely suggests that, at $\beta_c$, $E_3$ nullifies for finite but large $N$. However, Eq.~\eqref{hoff1} (and therefore Fig.~\ref{fig:E3}(a)) is inconclusive since it is not defined at $\beta_c$.

Therefore, in order to understand what really happens, we need to consider the exact, finite-$N$ solution of the 2D Ising model. This can be easily done using the transfer matrices \cite{Baxter}, and we made use of the ready formulas presented in \cite{Beale_1996, Izmailian_2002}. Feeding these formulas to \textsc{Mathematica}, we find the behaviour shown on Fig.~\ref{fig:E3}(a). Denoting maximum values of $|E_3|$ on both sides of $\beta_c$ as
\bea \label{pup1}
E_3^- := \max_{\beta < \beta_c} E_3 \qquad \mathrm{and} \qquad E_3^+ := \max\limits_{\beta > \beta_c} |E_3|,
\eea
we find that
\bea
E_3^{\pm} \propto N^{3/2},
\eea
and the distance of the peaks from $\beta_c$,
\bea \nonumber
\eta_- := \beta_c - \arg\max\limits_{\beta < \beta_c} E_3 \qquad \text{and} \qquad \eta_+ := \arg\max\limits_{\beta > \beta_c} |E_3| - \beta_c,
\eea
both scale as $N^{-1/2}$:
\bea \label{pup2}
\eta_{\pm} \propto N^{-1/2}.
\eea
Importantly, there is a certain assymmetry between the peaks in that they have slightly differing magnitudes and distances from $\beta_c$:
\bea \label{pup3}
0 &<& E_3^+ - E_3^- \propto N \ln N,
\\
0 &<& \eta_- - \eta_+ \propto N^{-1}.
\eea
This suggests that $E_3$ is not necessarily $0$ at $\beta_c$, which would of course not be surprising as $E_3$ is not zero even far away from criticality [see Fig.~\ref{fig:E3}(a)]. At most, $E_3(\beta_c)$ may scale proportionally with the peaks, i.e., as $N^{3/2}$, therefore, we can write
\bea
E_3(\beta_c) = O(N^{3/2}).
\eea

Keeping in mind that, away from criticality, $E_3 \propto N$, for the asymmetry, which we quantify by
\bea \label{skewnk}
\mathcal{A} = \frac{\sqrt[3]{E_3}}{\sqrt{E_2}},
\eea
we find that
\beaa \label{skewnk1}
\mathcal{A}(\beta \neq \beta_c) &\propto N^{-1/6},
\\
\mathcal{A}(\beta_c) &\propto \ln^{-1/2} N.
\eeaa
In both cases, the asymmetry tends to zero in the thermodynamic limit.

Note that $E_3$ peaking near the critical point but not \textit{at} it fits very well into the general picture drawn above. Indeed, as the temperature approaches $\beta_c$, the typical subset of energy levels approaches the less dense region characterized by the increased specific heat [see Eq.~\eqref{qnem}, keeping in mind that $K_2$ is given by Eq.~\eqref{K2}]. At some point, part of the subset will be in the critical, ``sparse,'' zone, whereas the other part will be in the noncritical, ``dense,'' zone, which will necessarily make the distribution asymmetric. Then, as one gets even closer to the critical point, most of the typical subset will be contained in the ``sparse'' zone, thereby mitigating the asymmetry.

In order to assess the Gaussianity of the energy distribution even further, let us invoke the fact that a distribution is Gaussian if and only if its first and second cumulants ($\kappa_1$ and $\kappa_2$) are nonzero whereas all the cumulants starting from the third ($\kappa_{k \geq 3}$) are zero. Now, we know that $\kappa_1 = \av{E}_\beta$ and $\kappa_2 = E_2$ and $\kappa_3 = E_3$. Therefore, if we quantify the Gaussianity by the relative weight of a cumulant as compared to $\kappa_2$, i.e, by $\kappa_k^{1/k} / \kappa_2^{1/2}$, then Eq.~\eqref{skewnk1} for the asymmetry $\mathcal{A}$ already gives an answer for the third cumulant. As per the fourth cumulant,
\bea
\kappa_4 = E_4 - 3 E_2^2 = -\frac{\partial E_3}{\partial \beta},
\eea
where the second equality is due to Eq.~\eqref{iterat}, we find numerically by calculating $\partial E_3 / \partial \beta$ that
\bea
\kappa_4(\beta_c) \propto N^2,
\eea
which means that
\beaa \label{No4}
\left. \frac{|\kappa_4|^{1/4}}{\kappa_2^{1/2}} \right\vert_{\beta \neq \beta_c} &\propto N^{-1/4},
\\
\left. \frac{|\kappa_4|^{1/4}}{\kappa_2^{1/2}} \right\vert_{\beta = \beta_c} &\propto \ln^{-1/2} N,
\eeaa
meaning that also the fourth cumulant becomes asymptotically insignificant in the thermodynamic limit. This shows that, as was proved generally in subsubsection \ref{app:alphazero}, at criticality, the energy distribution of the the square-lattice $2$D Ising model does indeed tend to a Gaussian.

Lastly, we note from Eqs.~\eqref{skewnk1} and \eqref{No4} that the convergence to Gaussianity at criticality is logarithmic, i.e., much slower than the polynomial convergence away from criticality.

\end{document}